# On the Nature of the Sommerfeld-Brillouin Forerunners (or Precursors)

P. K. Jakobsen[†] and M. Mansuripur[‡]

[†]Department of Mathematics and Statistics, UIT The Arctic University of Norway, Tromsø, Norway
[‡]College of Optical Sciences, The University of Arizona, Tucson, Arizona, U.S.A.

**Abstract**. We present a brief overview of Sommerfeld's forerunner signal, which occurs when a monochromatic plane-wave (frequency $\omega = \omega_s$) suddenly arrives, at time $t = 0$ and at normal incidence, at the surface of a dispersive dielectric medium of refractive index $n(\omega)$. Deep inside the dielectric host at a distance $z_0$ from the surface, no signal arrives until $t = z_0/c$, where $c$ is the speed of light in vacuum. Immediately after this point in time, however, a weak but extremely high frequency signal is observed at $z = z_0$. This so-called Sommerfeld forerunner (or precursor) is highly chirped, meaning that its frequency, which is much greater than $\omega_s$ immediately after $t = z_0/c$, declines rapidly with the passage of time. The incident light with its characteristic frequency $\omega_s$ eventually arrives at $t \cong z_0/v_g$, where $v_g$ is the group velocity of the incident light inside the host medium—it is being assumed here that $\omega_s$ is outside the anomalous dispersion region of the host. Brillouin has identified a second forerunner that occupies the interval between the end of the Sommerfeld forerunner at $t \cong n(0)z_0/c$ and the beginning of the steady signal (i.e., that which has the incident frequency $\omega_s$) at $t = z_0/v_g$. This second forerunner, which is also weak and chirped, having a frequency that is well below $\omega_s$ at first, then grows rapidly in time to reach $\omega_s$, is commonly referred to as the Brillouin forerunner (or precursor). Given that the incident wave has a sudden start at $t = 0$, its frequency spectrum spans the entire range of frequencies from $-\infty$ to $\infty$. Consequently, the high-frequency first forerunner cannot be considered a superoscillation, nor can the low-frequency second forerunner be regarded as a suboscillation. The goal of the present paper is to extend the Sommerfeld-Brillouin theory of precursors to bandlimited incident signals, in an effort to determine the conditions under which these precursors would continue to exist, and to answer the question as to whether or not such precursors, upon arising from a bandlimited incident signal, constitute super- or suboscillations.

**1. Introduction**. When a light pulse with a well-defined leading edge enters a homogeneous, isotropic, and dispersive dielectric medium at $t = 0$, it gets distorted and attenuated as it makes its way through the medium; see Fig.1(a). At a large distance $z_0$ from the point of entry, the leading edge of the pulse arrives at time $t_0 = z_0/c$, where $c$ is the speed of light in vacuum. Figure 1(b), adapted from Léon Brillouin's book on the subject,[1] shows that, in the vicinity of $t_0$, the light pulse has a weak amplitude but a high frequency; this is known as the first (or Sommerfeld) forerunner. A short while later, the character of the signal at $z = z_0$ changes; it now oscillates at a low frequency, albeit with a small amplitude, and is referred to as the second (or Brillouin) forerunner.[1,2] Eventually, the bulk of the signal arrives with the characteristic frequency of the incident beam. Sommerfeld has described the situation as follows:[3] "*If we let white light fall perpendicularly on a dispersive plate, then the less refracted (and hence 'faster') components of the white light do not precede the more refracted (and hence 'slower') components, and the light is not red at the first instance of emergence. Instead, the wave front of each component propagates with the same velocity c through the plate, and each component contributes equally to the energy of the initially emerging light. These initially emerging forerunners do not show the colors of the components of which they are composed; instead, they have an ultraviolet wavelength determined by the dispersive power and thickness of the plate, and a very small intensity.*"

The first forerunner thus appears to have the characteristics of a superoscillating signal, except that the incident pulse, due to its sharp leading-edge, is *not* a bandlimited waveform. In an effort to determine the circumstances under which the Sommerfeld forerunner could be considered a superoscillator (and the Brillouin forerunner a suboscillator), we extend the original Sommerfeld-Brillouin theory to bandlimited incident beams. The single-oscillator Lorentz model of the refractive index will be used throughout this paper, as was the case in the seminal publications of Sommerfeld and Brillouin.[3-5]

The forerunners (also called precursors) are discussed briefly in J.D. Jackson's *Classical Electrodynamics*, 2nd edition, pages 313-326, where he describes the behavior of both the first (Sommerfeld) and the second (Brillouin) precursors — Jackson's 3rd edition does *not* have all the interesting results.[2] According to Jackson, the first precursor's oscillation frequency at about $10^{20}$ Hz, is essentially independent of the incident light (visible range $\sim 5 \times 10^{14}$ Hz). This first-precursor's frequency is related to the properties of the host medium, and is associated with a saddle-point located on the far right-hand side of the complex $\omega$-plane (as well as its mirror image in the imaginary axis, located on the far left-hand side). The second (Brillouin) precursor occurs later in time, starting at $t \cong n(0)z_0/c$, where $n(0)$, the refractive index of the host medium at zero frequency ($\omega = 0$), is always greater than 1.0. The relatively low frequencies of the second precursor are associated with a pair of saddle-points located symmetrically with respect to the imaginary axis and near the origin of the complex $\omega$-plane.

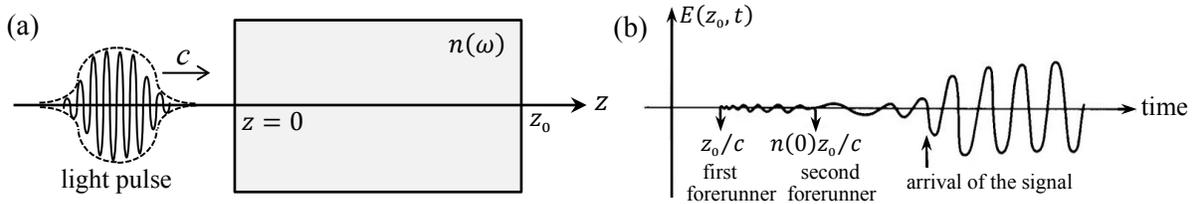

**Fig.1**. (a) A light pulse having a well-defined leading edge arrives at $t = 0$ at a dielectric slab of refractive index $n(\omega)$.
(b) Evolution of the light amplitude at $z = z_0$ following its initial arrival at $t_0 = z_0/c$ (adapted from [1]).

Jackson also uses the single Lorentz oscillator model for the host medium, and his incident beam is more or less monochromatic — albeit with the incident amplitude being zero prior to $t = 0$. As such, the phase refractive index $n$ of the medium is evaluated at the "monochromatic" frequency $\omega_s$ of the incident beam. Jackson places the peak of the incident spectrum at a fairly low frequency, namely, below the frequency of the real part of the pole of the Lorentz oscillator. In the case of the second precursor, a second pair of saddle-points, initially located on the imaginary axis but eventually moving out into the complex $\omega$-plane, becomes dominant, even though the contributions of the first pair of saddle-points (i.e., those on the far right- and far left-hand sides of the complex plane) to the observed signal at the far away point $z_0$ persist as weak oscillations superposed atop the (relatively stronger) Brillouin precursor.

The analysis presented in this paper aims to shed light on a somewhat broader problem of precursor oscillations. Starting with a bandlimited incident spectrum, we find that the high-frequency oscillations of the signal at the observation point coexist with the contributions from the end-points of the spectral bandwidth. The transition from the first (Sommerfeld) to the second (Brillouin) precursor is gradual, depending on the exact shape and location of the incident spectrum relative to the pole and zero of the Lorentz oscillator. Jackson's treatment, in particular, relies heavily on unwarranted approximations, as the authors of the old did not use numerics and, therefore, were forced to make rough approximations. By extending the prior work and focusing on bandlimited incident spectra (which Sommerfeld and Brillouin did not consider), we rely on a mixture of analytic and numeric techniques to shed new light on an old problem.

**2. Statement of the problem**. The electric-field amplitude at a distance $z = z_0$ from the entrance facet of the host medium is given by

$$E(z_0, t) = 2\text{Re}\left\{\int_{\omega_{\min}}^{\omega_{\max}} \mathcal{E}(\omega) \exp\{i\omega[n(\omega)(z_0/c) - t]\} d\omega\right\}$$

$$= 2\text{Re}\left\{\int_{\omega_{\min}}^{\omega_{\max}} \mathcal{E}(\omega) \exp\{i\zeta\omega[n(\omega) - \tau]\} d\omega\right\}. \tag{1}$$

Here, $\zeta = z_0/c$ and $\tau = ct/z_0$. The spectral profile $\mathcal{E}(\omega)$ of the incident $E$-field is band-limited, as it is confined to the interval $\omega_{\min} \leq |\omega| \leq \omega_{\max}$. For the $E$-field to be real-valued, the spectral distribution must be Hermitian, that is, $\mathcal{E}(-\omega) = \mathcal{E}^*(\omega)$. Although in the original Sommerfeld-Brillouin analysis the relevant range for the observation time was $\tau \geq 1$, no such restriction applies in the case of bandlimited incident beams. (In spite of the fact that nearly all the energy of a bandlimited wavepacket can be concentrated within a short pulse during a limited time interval, the packet will continue to have tails that extend to infinity both before and after the main body of the pulse. As such, there exists an observable signal at $z = z_0$ at all times $t$.)

In what follows, we shall employ the steepest-descent trajectories and, in particular, the method of saddle-point approximation,[6-8] to evaluate $E(z_0, t)$. To this end, it will be found convenient to write the complex entity $i\omega[n(\omega) - \tau]$ appearing in Eq.(1) as $\varphi(\omega) + i\psi(\omega)$.

The single-oscillator Lorentz model of classical electrodynamics describes the frequency-dependent refractive index $n(\omega)$ of the host medium in terms of a plasma frequency $\omega_p$, a resonance frequency $\omega_r$, and a damping coefficient $\gamma$, as follows:[2,9]

$$n(\omega) = \sqrt{1 + \frac{\omega_p^2}{\omega_r^2 - \omega^2 - i\gamma\omega}} = \sqrt{\frac{\omega^2 + i\gamma\omega - (\omega_r^2 + \omega_p^2)}{\omega^2 + i\gamma\omega - \omega_r^2}} = \sqrt{\frac{(\omega - \omega_a)(\omega + \omega_a^*)}{(\omega - \omega_b)(\omega + \omega_b^*)}}. \tag{2}$$

In this equation, $\omega_a = \sqrt{\omega_r^2 + \omega_p^2 - (\gamma/2)^2} - i(\gamma/2)$ and $\omega_b = \sqrt{\omega_r^2 - (\gamma/2)^2} - i(\gamma/2)$. Figure 2 shows the zeros (at $\omega = \omega_a$ and $-\omega_a^*$) and poles (at $\omega = \omega_b$ and $-\omega_b^*$) of the refractive index in the complex $\omega$-plane. The two short line-segments connecting each pole to its adjacent zero are branch-cuts.

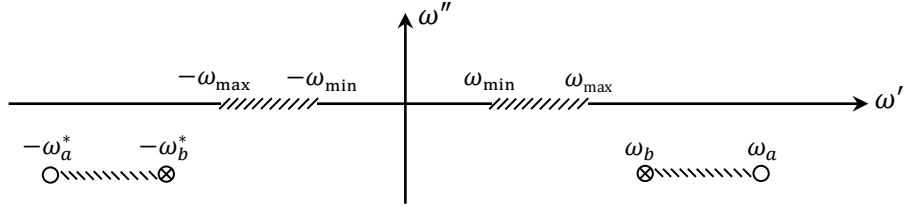

**Fig.2**. Complex-plane depiction of the zeros, $\omega_a, -\omega_a^*$, and poles, $\omega_b, -\omega_b^*$, of the refractive index $n(\omega)$ of the single-oscillator Lorentz model given in Eq.(2). The short line-segments connecting each pole to its adjacent zero are branch-cuts. The bandlimited spectrum $\mathcal{E}(\omega)$ of the incident beam is confined to the range $\omega_{\min} \leq |\omega| \leq \omega_{\max}$, and the incident spectrum is Hermitian, that is, $\mathcal{E}(-\omega) = \mathcal{E}^*(\omega)$.

The refractive index, being complex-valued in general, may be written in terms of its real and imaginary parts as $n(\omega) = n'(\omega) + in''(\omega)$. Typical plots of $n'(\omega)$ and $n''(\omega)$ are shown in Fig.3, with the frequency $\omega$ normalized by the (arbitrarily chosen) reference frequency $\omega_{\text{ref}} = 1.885 \times 10^{15}$ rad/sec ($\lambda_{\text{ref}} = 2\pi c/\omega_{\text{ref}} = 1.0$ μm). The branch-cut, shown in purple color slightly below the $\omega$-axis, extends from $\omega_b/\omega_{\text{ref}} \cong 4.0 - 0.05i$ to $\omega_a/\omega_{\text{ref}} \cong 14.56 - 0.05i$. Note that, in the vicinity of the branch-cut, $n'(\omega)$ is fairly small (albeit positive), whereas $n''(\omega)$ is quite large; this is the region where the dielectric medium shows strong absorption. For $\omega < \omega_b'$, absorption is relatively weak and the real part $n'(\omega)$ of the refractive index is greater than 1.0, which indicates that the phase velocity of propagation inside the medium is less than $c$. For $\omega > \omega_a'$, again absorption is weak, but $n'(\omega)$ is less than 1.0, which means that the phase velocity of propagation is greater than $c$.

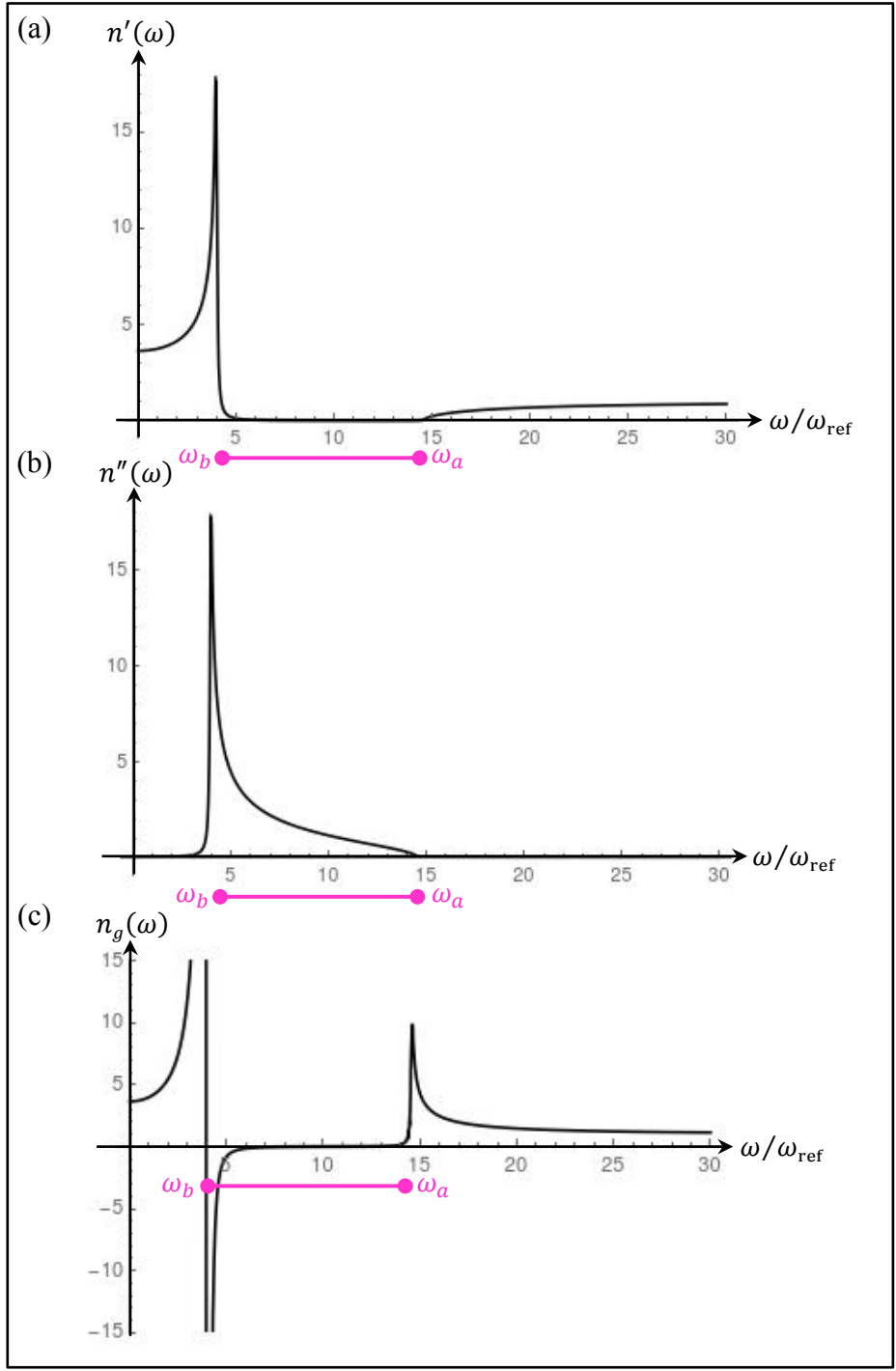

**Fig.3**. Host medium's refractive index $n(\omega)$ versus the frequency $\omega$ for a single-oscillator Lorentz model having $\omega_p = 14\omega_{\text{ref}}$, $\omega_r = 4\omega_{\text{ref}}$, $\gamma = 0.1\omega_{\text{ref}}$ ($\omega_{\text{ref}} = 1.885 \times 10^{15}$ rad/sec, corresponding to $\lambda_{\text{ref}} = 1.0$ μm, is a reference frequency chosen arbitrarily for normalization purposes). (a) The real part $n'(\omega)$ of the refractive index. (b) The imaginary part $n''(\omega)$ of the refractive index. (c) The group refractive index $n_g(\omega) = d[\omega n'(\omega)]/d\omega$. The group velocity $v_g(\omega) = c/n_g(\omega)$ approaches $c$, the speed of light in vacuum, when $\omega \to \infty$.

Another relevant function for wavepackets propagating inside the dielectric medium is the so-called group refractive index, defined as $n_g(\omega) = d[\omega n'(\omega)]/d\omega$. For a narrowband signal outside the region of the branch-cut, the group velocity of propagation is $v_g(\omega) = c/n_g(\omega)$. Figure 3(c) shows the corresponding plot of $n_g$ versus the frequency $\omega$. In the vicinity of the branch-cut, the group index behaves abnormally and has no physical significance. However, outside this frequency range, the group index is always greater than 1.0, signifying that, for narrowband wavepackets whose frequency spectrum falls either below $\omega_b'$ or above $\omega_a'$, the group velocity of propagation does not exceed $c$.

Treating $n(\omega)$ as a function of the complex frequency $\omega = \omega' + i\omega''$, one finds other interesting properties of the complex refractive index. For instance, setting the derivative with respect to $\omega$ of $n(\omega)$ equal to zero yields the only point in the entire $\omega$-plane where the refractive index is locally flat. According to Eq.(2), this is the point where the derivative of the denominator under the radical vanishes, that is,

$$dn(\omega)/d\omega = 0 \quad \rightarrow \quad d(\omega_r^2 - \omega^2 - i\gamma\omega)/d\omega = 0 \quad \rightarrow \quad \omega = -i\gamma/2. \tag{3}$$

The refractive index is thus seen to be flat in the vicinity of $\omega = -i\gamma/2$, which is the point where the extended branch-cuts meet the $\omega''$-axis. Along the $\omega''$-axis, $n''(\omega) = 0$ everywhere, and $n'(\omega)$ has a single peak at $\omega'' = -\gamma/2$. Along any given ray, $n(\omega) \rightarrow 1 - \tfrac{1}{2}(\omega_p/\omega)^2$ when $|\omega| \rightarrow \infty$. Thus, when $\omega' \rightarrow \pm\infty$ along the real axis, $n(\omega) \rightarrow 1$ from below, whereas along the imaginary axis, when $\omega'' \rightarrow \pm\infty$, $n(\omega) \rightarrow 1$ from above.

**3. Sommerfeld's original formulation**. Sommerfeld[1,3] begins by assuming that the incident $E$-field is a uniform-amplitude, single-frequency sinusoid that starts at $t = 0$, namely,

$$E(z = 0, t) = \text{step}(t)\sin(\omega_s t). \tag{4}$$

The Fourier transform of $E(z = 0, t)$ is readily evaluated, as follows:

$$\mathcal{E}(\omega) = \lim_{\alpha \to 0}(2\pi)^{-1} \int_0^\infty \exp(-\alpha t)\sin(\omega_s t)\exp(i\omega t)\,dt$$

$$= (4\pi i)^{-1} \lim_{\alpha \to 0} \int_0^\infty \{\exp[-(\alpha - i\omega_s - i\omega)t] - \exp[-(\alpha + i\omega_s - i\omega)t]\}dt$$

$$= (4\pi i)^{-1} \lim_{\alpha \to 0}[(\alpha - i\omega_s - i\omega)^{-1} - (\alpha + i\omega_s - i\omega)^{-1}]$$

$$= (4\pi i)^{-1} \lim_{\alpha \to 0}\left[\frac{\alpha}{\alpha^2 + (\omega+\omega_s)^2} + \frac{i(\omega+\omega_s)}{\alpha^2 + (\omega+\omega_s)^2} - \frac{\alpha}{\alpha^2 + (\omega-\omega_s)^2} - \frac{i(\omega-\omega_s)}{\alpha^2 + (\omega-\omega_s)^2}\right]$$

$$= (4\pi i)^{-1}[\pi\delta(\omega+\omega_s) + i(\omega+\omega_s)^{-1} - \pi\delta(\omega-\omega_s) - i(\omega-\omega_s)^{-1}]$$

$$= \tfrac{i}{4}\delta(\omega-\omega_s) - \frac{1}{4\pi(\omega-\omega_s)} - \tfrac{i}{4}\delta(\omega+\omega_s) + \frac{1}{4\pi(\omega+\omega_s)}. \tag{5}$$

Inside the host medium of refractive index $n(\omega)$, at a distance $z = z_0$ from the entrance facet, the $E$-field amplitude will be

$$E(z_0, t) = \int_{-\infty}^{\infty} \mathcal{E}(\omega) \exp\{i\zeta\omega[n(\omega) - \tau]\}\,d\omega \quad \leftarrow \boxed{\zeta = z_0/c,\ \tau = ct/z_0}$$

$$= \tfrac{i}{4}\exp\{i\zeta\omega_s[n(\omega_s) - \tau]\} - \tfrac{i}{4}\exp\{-i\zeta\omega_s[n^*(\omega_s) - \tau]\}$$

$$-\frac{1}{4\pi}\int_{-\infty}^{\infty}\frac{\exp\{i\zeta\omega[n(\omega)-\tau]\}}{\omega-\omega_s}\,d\omega + \frac{1}{4\pi}\int_{-\infty}^{\infty}\frac{\exp\{i\zeta\omega[n(\omega)-\tau]\}}{\omega+\omega_s}\,d\omega$$

$$= \tfrac{1}{2}e^{-\omega_s n''(\omega_s)z_0/c} \sin\{\omega_s[t - n'(\omega_s)z_0/c]\} \quad \leftarrow \boxed{n(\omega) = n' + in''}$$

$$-(2\pi)^{-1}\mathrm{Re}\left\{\int_{-\infty}^{\infty} \frac{\exp\{i\zeta\omega[n(\omega)-\tau]\}}{\omega-\omega_s} d\omega\right\}. \tag{6}$$

The integral in Eq.(6) is a principal-value integral that must be evaluated over the real axis $\omega'$, first from $-\infty$ to $\omega_s - \varepsilon$, and then from $\omega_s + \varepsilon$ to $\infty$. If, instead, we choose a continuous path in the complex $\omega$-plane consisting of the slightly deformed real axis around a small semi-circle in the upper half-plane centered at $\omega = \omega_s$, as shown in Fig.4, then the semi-circle's contribution accounts for the first term on the right-hand side of Eq.(6). This simplifies the expression of $E(z_0, t)$ by reducing the right-hand side of Eq.(6) to just the integral term, with the integration contour now including a small semi-circle in the upper half-plane around $\omega = \omega_s$.

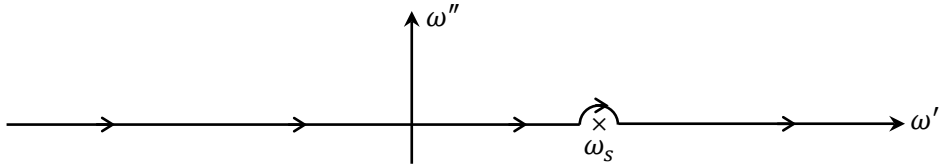

**Fig.4**. The contour of integration in the complex $\omega$-plane consists of the real axis $\omega'$ deformed around the small semi-circle above the pole that is associated with the source frequency $\omega_s$.

When $t < z_0/c$, the integration contour can be closed in the upper-half plane with an infinitely large semi-circle, resulting in $E(z_0, t) = 0$ for $\tau < 1$. When $\tau > 1$, one must check the integration path that replaces the original contour to see whether the closed path includes or excludes the pole at $\omega = \omega_s$. Once this pole is found to be included, its residue (times $2\pi i$) must be added to the final result of integration. This means that, for sufficiently large $t$, as expected, $E(z_0, t)$ will contain the term $\exp[-\omega_s n''(\omega_s) z_0/c] \sin\{\omega_s[t - n'(\omega_s) z_0/c]\}$.

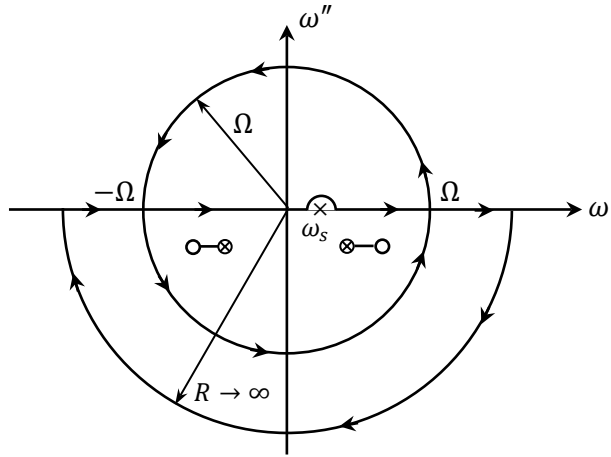

**Fig.5**. The contour of integration on the real axis is deformed into a circle of radius $\Omega$. The contribution to the integral by the segment between $-\Omega$ and $\Omega$ (including the small semicircle surrounding the source frequency $\omega_s$) is replaced by the semicircle of radius $\Omega$ in the upper half-plane. The remaining parts of the real axis, namely the segments $(-\infty, -\Omega)$ and $(\Omega, \infty)$, are replaced by the lower semicircle of radius $\Omega$. The latter replacement is justified by the fact that the integral over the semi-annular contour in the lower half-plane vanishes, and that, for $\tau > 1$, the integral over the infinitely large semicircle of radius $R$ goes to zero.

Sommerfeld gives a simple approximate formula for the first forerunner, which he obtains as follows. Suppose the integration contour of Fig.4 is replaced by the circular contour of large (but not infinite) radius $\Omega$ shown in Fig.5, where $\omega = \Omega e^{i\theta}$ and $0 \le \theta \le 2\pi$. On this circle, $n(\omega) \cong 1 + \frac{1}{2}\omega_p^2/(\omega_r^2 - \omega^2 - i\gamma\omega) \cong 1 - \frac{1}{2}(\omega_p^2/\omega^2)$. We now write

$$E(z_o, t) = -(2\pi)^{-1} \text{Re}\left\{\int_{-\infty}^{\infty} \frac{\exp\{i\zeta\omega[n(\omega)-\tau]\}}{\omega - \omega_s} d\omega\right\}$$

$$\cong (2\pi)^{-1} \text{Re}\left\{\int_0^{2\pi} \frac{\exp\{i\zeta\omega[1-\tau-\frac{1}{2}(\omega_p^2/\omega^2)]\}}{\omega[1-(\omega_s/\omega)]} i\Omega e^{i\theta} d\theta\right\}$$

$$\cong (2\pi)^{-1} \text{Re}\left\{i\int_0^{2\pi} \left(1 + \frac{\omega_s}{\omega}\right)\exp\{-i[(t-\zeta)\Omega e^{i\theta} + (\zeta\omega_p^2/2\Omega)e^{-i\theta}]\} d\theta\right\}. \quad (7)$$

Choosing $\Omega$ such that $(t-\zeta)\Omega = \zeta\omega_p^2/2\Omega$, that is,

$$\Omega = \omega_p\sqrt{\zeta/2(t-\zeta)} = \omega_p\sqrt{z_o/2(ct-z_o)}, \quad (8)$$

we will have

$$E(z_o, t) \cong (2\pi)^{-1}\text{Re}\left\{i\int_0^{2\pi}\left[1 + \frac{\omega_s \exp(-i\theta)}{\omega_p\sqrt{z_o/2(ct-z_o)}}\right]\exp[-i\omega_p\sqrt{(2z_o/c)(t-z_o/c)}\cos\theta] d\theta\right\}$$

$$= (2\pi)^{-1}\int_0^{2\pi} \sin[\omega_p\sqrt{(2z_o/c)(t-z_o/c)}\cos\theta] d\theta \;\longleftarrow\; \boxed{\text{odd symmetry around } \theta = \pi/2, 3\pi/2}$$

$$+ \left(\frac{\omega_s\sqrt{t-z_o/c}}{\pi\omega_p\sqrt{2z_o/c}}\right)\text{Re}\left\{i\int_0^{2\pi}\exp[-i\theta - i\omega_p\sqrt{(2z_o/c)(t-z_o/c)}\cos\theta] d\theta\right\}$$

$$= \left(\frac{2\omega_s\sqrt{t-z_o/c}}{\omega_p\sqrt{2z_o/c}}\right) J_1[\omega_p\sqrt{(2z_o/c)(t-z_o/c)}]. \;\longleftarrow\; \boxed{J_1(\cdot) \text{ is Bessel function of 1}^{st}\text{ kind, order 1}} \quad (9)$$

Invoking the large-argument asymptotic approximation $J_1(x) \sim -\sqrt{2/(\pi x)}\cos(x + \frac{1}{4}\pi)$, we now obtain

$$E(z_o, t) \cong -\frac{\omega_s}{\omega_p\sqrt{\pi\omega_p}}\left(\frac{2c}{z_o}\right)^{3/4}(t - z_o/c)^{1/4}\cos[\omega_p\sqrt{(2z_o/c)(t-z_o/c)} + \frac{1}{4}\pi]. \quad (10)$$

The Sommerfeld precursor's oscillation frequency $\omega_{\text{spc}}$, being the time derivative of the argument of the cosine function in Eq.(10), is readily found to be

$$\omega_{\text{spc}}(t) = \frac{\omega_p}{\sqrt{2[(ct/z_o)-1]}} = \frac{\omega_p}{\sqrt{2(\tau-1)}} = \omega_{\text{saddle}}. \;\longleftarrow\; \boxed{\text{Saddle on the right-hand side of } \omega\text{-plane; see Sec.6, and also Appendix C, Eq.(C1).}} \quad (11)$$

The above equation now yields

$$t - z_o/c = (z_o/2c)(\omega_p/\omega_{\text{spc}})^2. \quad (12)$$

Brillouin[1,4] gives a correction to Sommerfeld's precursor oscillation amplitude $E_{\text{spc}}$ of Eq.(10) due to absorption within the host medium, which Sommerfeld had originally ignored. Brillouin's correction term is the exponential decay factor $\exp[-\gamma(t - z_o/c)]$, resulting in the following precursor amplitude (see Appendix A for a derivation of the decay factor):

$$E_{\text{spc}}(t) \cong \omega_s (2c/z_0)^{3/4} (t - z_0/c)^{1/4} e^{-\gamma(t-z_0/c)} / \sqrt{\pi \omega_p^3}$$
$$= (\omega_s/\omega_p)\sqrt{2c/(\pi z_0 \omega_{\text{spc}})} \exp[-\tfrac{1}{2}\gamma(\omega_p/\omega_{\text{spc}})^2 (z_0/c)]. \tag{13}$$

Brillouin's numerical results are based on the following set of parameters: $\omega_s = 4 \times 10^{15}$, $\lambda_s = 2\pi c/\omega_s = 0.47$ μm, $\omega_p = \omega_r = 10\omega_s$, $\gamma = 0.15\omega_r$, and $z_0 = 1.0$ cm, which yield

$$n(\omega_s) \cong \sqrt{2}[1 + \tfrac{1}{4}(\omega_s/\omega_r)^2 + i(\tfrac{1}{4}\gamma\omega_s/\omega_r^2)] \cong 1.418 + 0.0053i, \tag{14}$$

and, for $\omega \gg \omega_s$,

$$n(\omega) \cong [1 - 50(\omega_s/\omega)^2] + i50\gamma(\omega_s^2/\omega^3) \cong 1 + i75(\omega_s/\omega)^3. \tag{15}$$

With these parameters, Eq.(13) becomes

$$E_{\text{spc}}(t) = 0.0342(t - \zeta)^{1/4} e^{-\gamma(t-\zeta)} = \left(\frac{13820}{\sqrt{\omega_{\text{spc}}}}\right) e^{-50\gamma(\omega_s/\omega_{\text{spc}})^2 \zeta}. \quad \leftarrow \boxed{\zeta = z_0/c} \tag{16}$$

The above amplitude peaks at $t_{\max} - \zeta = 1/(4\gamma)$, at a maximum of $E_{\text{spc}}(t_{\max}) \cong 2.14 \times 10^{-6}$. Compare this result with the incident $E$-field amplitude at a precursor frequency $\omega \gg \omega_s$, namely, $\mathcal{E}(\omega) = 1/2\pi(\omega - \omega_s)$, which, after attenuation due to travel through the host medium, arrives at $z = z_0$ with the following amplitude:

$$\text{attenuated } \mathcal{E}(\omega) = \frac{\exp[-\omega n''(\omega) z_0/c]}{2\pi(\omega - \omega_s)} = \frac{\exp[-50\gamma(\omega_s/\omega)^2 \zeta]}{2\pi(\omega - \omega_s)}. \tag{17}$$

A comparison of Eqs.(16) and (17) reveals that, during the short time interval when the Sommerfeld precursor at $z = z_0$ has a frequency $\omega \cong \omega_{\text{spc}} \gg \omega_s$, its amplitude is far greater than the corresponding spectral amplitude by the enormous factor of $\sim 8.7 \times 10^4 \sqrt{\omega_{\text{spc}}}$. This, however, is an unfair comparison, considering that the entire spectral content of the incident beam at this frequency is squeezed into a short time interval, then embedded within the precursor signal. The proper comparison must be made between the energy content of the precursor signal on the one hand, and the spectral energy of the incident beam, on the other, as we now show.

The energy-density associated with the incident spectral distribution at frequency $\omega$ is obtained from Eq.(5), as follows:

$$\mathcal{E}^2(\omega) = \left(\frac{1}{4\pi}\right)^2 \left(\frac{1}{\omega + \omega_s} - \frac{1}{\omega - \omega_s}\right)^2 = \frac{1}{16\pi^2}\left[\frac{1}{(\omega + \omega_s)^2} + \frac{1}{(\omega - \omega_s)^2} + \frac{1}{\omega_s(\omega + \omega_s)} - \frac{1}{\omega_s(\omega - \omega_s)}\right]. \tag{18}$$

To evaluate the energy content of the incident beam in the entire frequency range $|\omega| \geq \omega_0$, where $\omega_0 \gg \omega_s$ is a large (but otherwise arbitrary) frequency, we write

$$\int_{-\infty}^{-\omega_0} \mathcal{E}^2(\omega) d\omega + \int_{\omega_0}^{\infty} \mathcal{E}^2(\omega) d\omega = \frac{1}{8\pi^2}\left[\int_{\omega_0}^{\infty} \frac{d\omega}{(\omega + \omega_s)^2} + \int_{\omega_0}^{\infty} \frac{d\omega}{(\omega - \omega_s)^2} + \frac{1}{\omega_s}\int_{\omega_0}^{\infty} \frac{d\omega}{\omega + \omega_s} - \frac{1}{\omega_s}\int_{\omega_0}^{\infty} \frac{d\omega}{\omega - \omega_s}\right]$$

$$= \frac{1}{8\pi^2}\left[\frac{1}{\omega_0 + \omega_s} + \frac{1}{\omega_0 - \omega_s} + \frac{1}{\omega_s} \ln\left(\frac{\omega_0 - \omega_s}{\omega_0 + \omega_s}\right)\right]. \tag{19}$$

Using a Taylor series expansion of its various terms, Eq.(19) is rewritten as follows:

Energy content in $|\omega| \geq \omega_0 = \frac{1}{8\pi^2}\left[\frac{1}{\omega_0}\sum_{k=0}^{\infty}(-\omega_s/\omega_0)^k + \frac{1}{\omega_0}\sum_{k=0}^{\infty}(\omega_s/\omega_0)^k \right.$ $\leftarrow \boxed{(1+x)^{-1} = \sum_{k=0}^{\infty}(-x)^k}$

$\boxed{\ln(1+x) = -\sum_{k=1}^{\infty}(-x)^k/k} \rightarrow \left. -\frac{1}{\omega_s}\sum_{k=1}^{\infty}(\omega_s/\omega_0)^k/k + \frac{1}{\omega_s}\sum_{k=1}^{\infty}(-\omega_s/\omega_0)^k/k\right]$

$$= \tfrac{1}{4\pi^2}\left\{\tfrac{1}{\omega_0}\sum_{k=0}^{\infty}(\omega_s/\omega_0)^{2k} - \tfrac{1}{\omega_s}\sum_{k=1}^{\infty}[(\omega_s/\omega_0)^{2k-1}/(2k-1)]\right\}$$

$$= \tfrac{1}{4\pi^2}\left\{\sum_{k=1}^{\infty}(\omega_s^{2k}/\omega_0^{2k+1}) - \sum_{k=1}^{\infty}[(\omega_s^{2k}/\omega_0^{2k+1})/(2k+1)]\right\}$$

$$= \tfrac{1}{2\pi^2\omega_0}\sum_{k=1}^{\infty}[k/(2k+1)](\omega_s/\omega_0)^{2k}$$

$$= \tfrac{\omega_s^2}{6\pi^2\omega_0^3}\left\{1 + \sum_{k=1}^{\infty}\left[\tfrac{k+1}{(2k/3)+1}\right]\left(\tfrac{\omega_s}{\omega_0}\right)^{2k}\right\}. \tag{20}$$

Since, by assumption, $\omega_0 \gg \omega_s$, the high-order terms in Eq.(20) are negligible, and the overall energy in the tails of the incident spectrum is approximated as $\sim \omega_s^2/(6\pi^2\omega_0^3)$. Returning to the Sommerfeld precursor oscillation amplitude given by Eq.(13)—and considering that the exponential decay factor accounts solely for absorption within the dielectric host—one may omit the decay factor and write the average intensity (i.e., half the squared amplitude) of the sinusoidal precursor as

$$\tfrac{1}{2}E_{\text{spc}}^2(t) = \omega_s^2(2c/z_0)^{3/2}(t - z_0/c)^{1/2}/(2\pi\omega_p^3). \tag{21}$$

Integrating the above intensity from $t = z_0/c$ to $t = t_0$ now yields

$$\tfrac{1}{2}\int_{t=z_0/c}^{t_0}E_{\text{spc}}^2(t)\mathrm{d}t = \omega_s^2(2c/z_0)^{3/2}(t_0 - z_0/c)^{3/2}/(3\pi\omega_p^3). \tag{22}$$

According to Eq.(12), $t_0 - z_0/c$ equals $(z_0/2c)(\omega_p/\omega_{\text{spc}})^2$, which is now substituted in Eq.(22) to yield the precursor's cumulative energy up to $t = t_0$ as $\omega_s^2/(3\pi\omega_{\text{spc}}^3)$. As the brief digression below demonstrates, Parseval's theorem requires the cumulative energy to be divided by $2\pi$ in order to compare with the spectral energy of Eq.(20). It is thus seen that the energy content of the precursor signal between $t = z_0/c$, where the oscillation frequency is infinite, and $t = t_0$, where the oscillation frequency is $\omega_{\text{spc}}$, is the same as the leading (and dominant) term in Eq.(20), provided that $\omega_0$ is equated with $\omega_{\text{spc}}$.

---

**Digression**: For the Fourier transform formula used in this paper, Parseval's theorem is derived below.

$$\int_{-\infty}^{\infty}F(\omega)G^*(\omega)\mathrm{d}\omega = \int_{-\infty}^{\infty}\left[\tfrac{1}{2\pi}\int_{-\infty}^{\infty}f(t)e^{i\omega t}\mathrm{d}t\right]G^*(\omega)\mathrm{d}\omega = \tfrac{1}{2\pi}\int_{-\infty}^{\infty}f(t)[\int_{-\infty}^{\infty}G(\omega)e^{-i\omega t}\mathrm{d}\omega]^*\mathrm{d}t$$

$$= (2\pi)^{-1}\int_{-\infty}^{\infty}f(t)g^*(t)\mathrm{d}t.$$

---

The general picture emerging from the above analysis by Sommerfeld (and its improvement as well as extension by Brillouin) is as follows. At the observation point $z = z_0$, the signal should be zero for $t < z_0/c$. Also, for $t > z_0/v_g(\omega_s)$—where $v_g$ is the group velocity at the frequency $\omega_s$ of the incident signal—the observed waveform should have more or less the same frequency as the incident light. This leaves the very narrow time window $z_0/c < t < z_0/v_g(\omega_s)$ for all the other frequencies to be squeezed in. These other frequencies arrive in rapid succession depending on their group velocity, which is only slightly less than $c$ at high frequencies, but continually declines as the frequency sweeps lower. Once the sweep of the high frequencies end, we arrive at the end of the first (Sommerfeld) precursor. At this point, the low frequencies (i.e., those below $\omega_s$) begin to arrive, and this is the beginning of the second (Brillouin) precursor. The group velocity for these low frequencies is less than that for the high frequencies, but higher than $v_g(\omega_s)$. Once all the low frequencies arrive, the entire spectrum (except for the part near $\omega_s$) has been exhausted; this point marks the end of the second (Brillouin) precursor. Beyond this point in time, the main signal frequency $\omega_s$ arrives, which will then persist as $t \to \infty$.

The above picture leaves no room for the interpretation of the observed precursors as "leaky" super- or suboscillations.[10] Nevertheless, it is a veritable picture that shows how the dispersive propagation through a dielectric slab can "sort out" the various frequencies in accordance with their (local) group velocities.

**4. Frequency spectra of finite-bandwidth incident beams**. For a finite-bandwidth incident light pulse, we define the center of the incident spectrum as $\omega_c = \frac{1}{2}(\omega_{min} + \omega_{max})$, and the incident bandwidth as $\Delta\omega = \omega_{max} - \omega_{min}$. The simplest incident spectrum will be uniform across the entire bandwidth, that is,

$$\mathcal{E}_1(\omega) = \text{rect}[(\omega - \omega_c)/\Delta\omega]; \qquad (\omega_{min} \leq |\omega| \leq \omega_{max}). \tag{23}$$

This spectrum and its inverse Fourier transform, i.e., the corresponding incident light pulse as a function of time at $z = 0$, are shown in Fig.6(a).

Another possible spectrum is one that peaks at $\omega_c$, then gently drops to zero at the edges of the bandwidth. A prototypical example is

$$\mathcal{E}_2(\omega) = \cos^2[\pi(\omega - \omega_c)/\Delta\omega]; \qquad (\omega_{min} \leq |\omega| \leq \omega_{max}). \tag{24}$$

Note that the requirement $\mathcal{E}(-\omega) = \mathcal{E}^*(\omega)$ demands in this case that $\omega_c/\Delta\omega$ be an integer. Also, it is not necessary for the power of the cosine function to be 2; any even integer will be acceptable. A variation on the same theme would be the following (bandlimited) spectrum:

$$\mathcal{E}_3(\omega) = \sin^{100}[\pi\omega/(2\omega_s)], \qquad |\omega| \leq 2\omega_s. \tag{25}$$

Plots of $\mathcal{E}_3(\omega)$ and its corresponding wavepacket as a function of time at $z = 0$ are shown in Fig.6(b). Although the tails of the wavepacket appear to be quite weak in this example, the fact that the spectrum is confined to the finite interval $(-2\omega_s, 2\omega_s)$ ensures that the tails extend all the way to infinity along the time axis.

A fourth possible spectrum also peaks somewhere between $\omega_{min}$ and $\omega_{max}$, but drops smoothly to zero at the edges $\omega_{min}$ and $\omega_{max}$ of the bandwidth, in such a way as to keep the spectrum differentiable (with respect to $\omega$) across the entire bandwidth $(\omega_{min}, \omega_{max})$. Thus,

$$\mathcal{E}_4(\omega) = \exp\left[\frac{1}{(\omega^2 - \omega_{min}^2)(\omega^2 - \omega_{max}^2)}\right]; \qquad (\omega_{min} \leq |\omega| \leq \omega_{max}). \tag{26}$$

The peak of the above spectrum occurs at $\omega_{peak} = \pm\sqrt{\frac{1}{2}(\omega_{min}^2 + \omega_{max}^2)}$, where $\mathcal{E}_4(\omega_{peak}) = \exp[-1/(\omega_c\Delta\omega)^2]$. Figure 6(c) shows plots of $\mathcal{E}_4(\omega)$ and its corresponding wavepacket. As before, the tails of the wavepacket persist all the way to infinity, although the rate of decline of the tail amplitude is quite rapid as $|t| \to \infty$. Note that, in the complex $\omega$-plane, $\mathcal{E}_4(\omega)$ has singularities at $\omega = \pm\omega_{min}$ and $\omega = \pm\omega_{max}$, which must be circumvented by going around them on circular arcs of vanishing radii. (Whereas the arc around $\omega_{min}$ must lie on the right-half of a small circle, that around $\omega_{max}$ must lie on the left-half of a corresponding circle.) Note also that $\mathcal{E}_4(\omega) \to 1$ when $|\omega| \to \infty$.

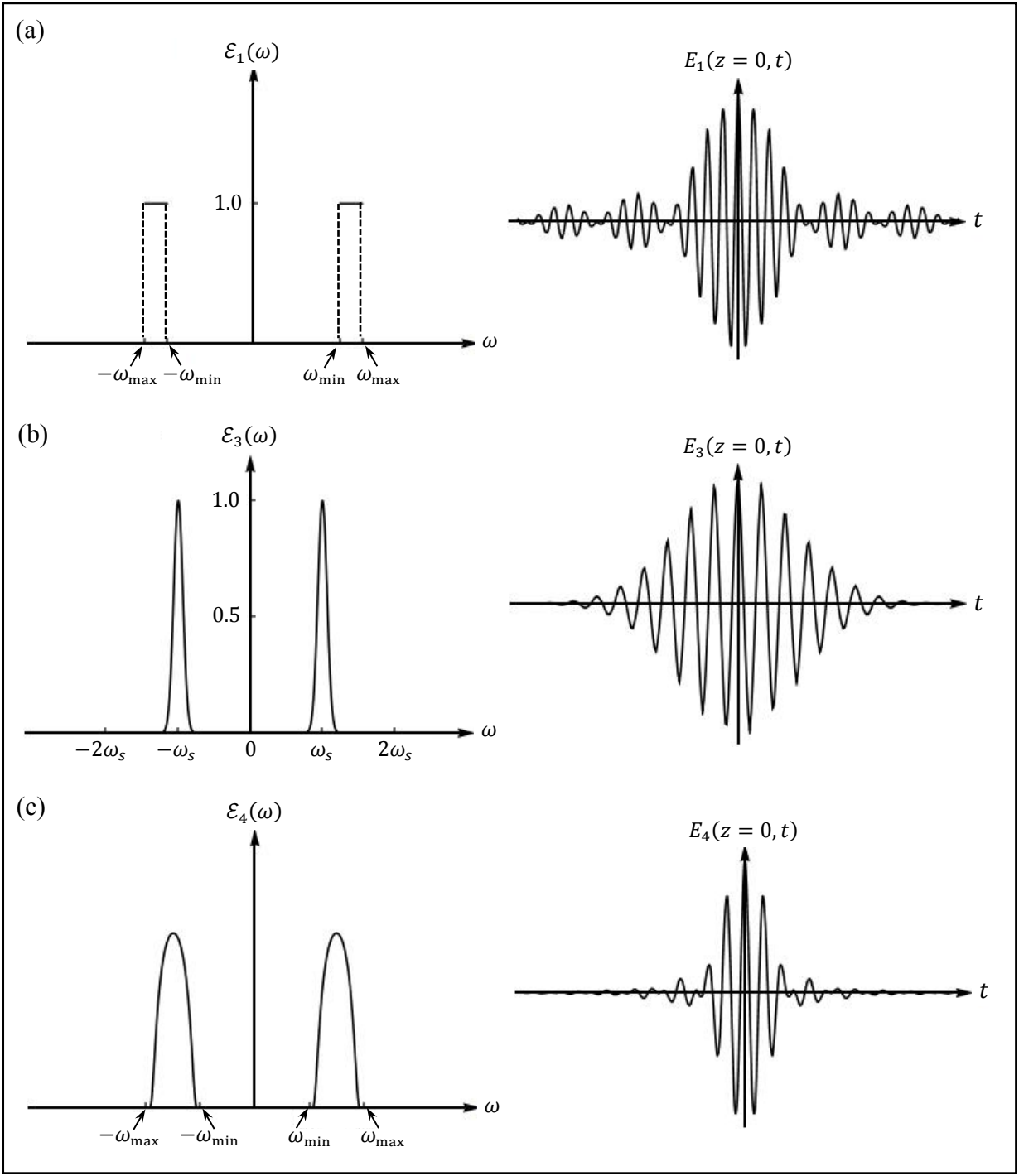

**Fig.6**. Examples of bandlimited spectra and their corresponding wavepackets in the time domain. (a) $\mathcal{E}(\omega) = 1.0$ when $\omega_{min} \leq |\omega| \leq \omega_{max}$. (b) $\mathcal{E}(\omega) = \sin^{100}[\pi\omega/(2\omega_s)]$ when $|\omega| \leq 2\omega_s$. (c) $\mathcal{E}(\omega) = \exp[(\omega^2 - \omega_{min}^2)^{-1}(\omega^2 - \omega_{max}^2)^{-1}]$ when $\omega_{min} \leq |\omega| \leq \omega_{max}$.

Other variations on the same theme as $\mathcal{E}_4(\omega)$ are also possible. For instance, if the goal is to shift the bulk of the spectrum away from the center $\omega_c$ of the $(\omega_{min}, \omega_{max})$ interval, one may resort to the following spectral distribution:

$$\mathcal{E}_5(\omega) = \exp\left[-\frac{\omega^{2p}}{(\omega^2 - \omega_{min}^2)^m (\omega_{max}^2 - \omega^2)^n}\right]. \tag{27}$$

Here, the $p, m, n$ parameters are positive integers, with $p \leq m + n$. The function $\mathcal{E}_5(\omega)$ peaks at a point between $\omega_{min}$ and $\omega_{max}$, which can be placed arbitrarily close to $\omega_{min}$ (or $\omega_{max}$) by a proper choice of the positive integers $p$, $m$, and $n$. In the special case of $p = m + n$, for example, it is easy to show that $\omega_{peak} = \pm\sqrt{(m+n)\omega_{min}^2\omega_{max}^2/(m\omega_{min}^2 + n\omega_{max}^2)}$. Note that, in the limit when $|\omega| \to \infty$, $\mathcal{E}_5(\omega) \to 1$ provided that $p < m + n$. For $p = m + n$, when $|\omega| \to \infty$, we find that $\mathcal{E}_5(\omega)$ goes to $e^{-1}$ if $n$ is even, and to $e$ if $n$ is odd.

The general goal is to approximate the integral in Eq.(1) by modifying the integration path, i.e., replacing the straight line-segment along the real axis that connects $\omega_{min}$ to $\omega_{max}$ with a combination of steepest-descent and/or stationary-phase contours that depart from the real axis at $\omega_{min}$, connect with other contours (preferably those that go through saddle points of the integrand), then return to the real axis at $\omega_{max}$; a typical example appears in Fig.7. Contacts among the steepest-descent and/or stationary-phase contours may occur at various locations in the complex $\omega'\omega''$-plane, including the singularities of the integrand, or points located at infinity. Throughout the following discussion, we shall be guided by the Cauchy-Goursat theorem of complex analysis, by the properties of meromorphic functions, and by the principles of steepest-descent and stationary-phase approximation.[6-8]

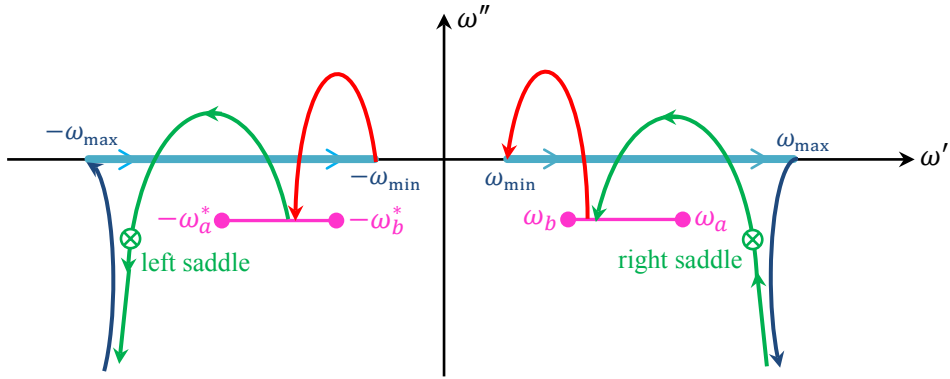

**Fig.7**. For a bandlimited signal whose spectrum is confined to the interval $\omega_{min} \leq |\omega| \leq \omega_{max}$, the contour of integration is deformed as follows: From $\omega_{max}$ we move out along a steepest-descent path to infinity in the lower-half of the complex $\omega'\omega''$-plane. Then, from $-\infty$, we return along the steepest-descent path that goes through the saddle point on the right-hand side; this path terminates on the branch-cut. Subsequently, from a nearby point on the branch-cut, we move along another steepest-descent path that brings us to $\omega_{min}$. These three segments of the deformed integration contour appear as blue, green, and red, respectively. A similar path exists on the left-half of the complex plane, replacing the straight-line integration contour from $-\omega_{max}$ to $-\omega_{min}$.

It must be pointed out that, in the region immediately above the branch-cut on the right-hand side of Fig.7, the refractive index $n(\omega)$ is imaginary and positive, whereas in the corresponding region on the left-hand side, $n(\omega)$ is imaginary and negative. Thus, along the immediate upper sides of both these branch-cuts, the real part $\varphi(\omega)$ of the exponent of the integrand in Eq.(1) will be negative. All steepest-descent trajectories that land on these branch-cuts from above thus arrive at locations where the integrand is quite small. This makes the contributions to the overall integral of those short segments of the branch-cuts that connect two adjacent contours fairly small and, therefore, negligible.

**5. Steepest-descent and stationary-phase trajectories**. The exponent of the integrand in Eq.(1), aside from the large coefficient $\zeta = z_0/c$, can be broken up into its real and imaginary parts, as follows:

$$\varphi(\omega) + i\psi(\omega) = i\omega[n(\omega) - \tau] \quad \leftarrow \boxed{\tau = ct/z_0}$$
$$= -\{[n'(\omega) - \tau]\omega'' + \omega' n''(\omega)\} + i\{[n'(\omega) - \tau]\omega' - \omega'' n''(\omega)\}. \quad (28)$$

The contours of steepest-descent are curves in the $\omega$-plane on which $\psi(\omega) = \psi_0$, where $\psi_0$ is an arbitrary (real-valued) constant. These curves typically start and stop at infinity, or at one of the poles $\omega_b, -\omega_b^*$, or on the branch-cuts. At the zeros $-\omega_a^*$ and $\omega_a$ of the refractive index, where $n' = n'' = 0$, we have $\varphi = \tau\omega_a'' = -\tau\gamma/2$ and $\psi = \pm\tau\omega_a' = \pm\tau\sqrt{\omega_r^2 + \omega_p^2 - (\gamma/2)^2}$.

Considering the special symmetry of $n(\omega)$ relative to the imaginary axis of the $\omega$-plane, namely, $n(-\omega^*) = n^*(\omega)$, it is readily seen that $\varphi(-\omega^*) = \varphi(\omega)$ and $\psi(-\omega^*) = -\psi(\omega)$.

When $\omega \to \infty$ along any ray in the complex plane, $n(\omega)$ approaches $1 - \tfrac{1}{2}(\omega_p/\omega)^2$, in which case the exponent of the integrand in Eq.(1) approaches a certain limit, as follows:

$$i\omega[n(\omega) - \tau] \to i(\omega' + i\omega'')[1 - \tfrac{1}{2}(\omega_p/|\omega|e^{i\theta})^2 - \tau]$$
$$= \{(\tau - 1)\omega'' + \tfrac{1}{2}(\omega_p/|\omega|)^2[\omega'' \cos(2\theta) - \omega' \sin(2\theta)]\}$$
$$-i\{(\tau - 1)\omega' + \tfrac{1}{2}(\omega_p/|\omega|)^2[\omega'' \sin(2\theta) + \omega' \cos(2\theta)]\}. \quad (29)$$

For any contour of constant-phase (i.e., steepest-descent trajectory), there is a specific value of $\omega'$ to which the contour asymptotically approaches as $|\omega| \to \infty$. The contour typically approaches a large positive $\omega'$ from the left-hand side, or a large negative $\omega'$ from the right-hand side, and eventually becomes parallel to the imaginary axis. Similarly, for any contour of constant-amplitude (i.e., stationary-phase trajectory), there exists a specific value of $\omega''$ to which the contour asymptotically approaches as $|\omega| \to \infty$. The contour typically approaches a negative $\omega''$ from above, and eventually parallels the real axis.

Next, we examine the immediate neighborhood of the zero $\omega = \omega_a$ of $n(\omega)$, where

$$i\omega[n(\omega) - \tau] \to i\omega_a\left[\frac{(\omega_a + \omega_a^*)^{1/2}(\omega - \omega_a)^{1/2}}{(\omega_a - \omega_b)^{1/2}(\omega_a + \omega_b^*)^{1/2}} - \tau\right]. \quad (30)$$

Let $\omega - \omega_a = \varepsilon \exp(i\theta)$, where $|\varepsilon| \ll 1$, and $0 \le \theta < 2\pi$ is measured counterclockwise from the positive real axis $\omega'$. Equation (30) may be further simplified, as follows:

$$i\omega[n(\omega) - \tau] \to (-\omega_a'' + i\omega_a')\left[\frac{(2\omega_a')^{1/2}\sqrt{\varepsilon}\exp(i\theta/2)}{(\omega_a' - \omega_b')^{1/2}(\omega_a' + \omega_b')^{1/2}} - \tau\right]$$
$$= (\tfrac{1}{2}\gamma + i\omega_a')\left\{\sqrt{2\omega_a'\varepsilon/(\omega_a'^2 - \omega_b'^2)}[\cos(\theta/2) + i\sin(\theta/2)] - \tau\right\}$$
$$= \tfrac{1}{2}\{-\gamma\tau + (1/\omega_p)\sqrt{2\omega_a'\varepsilon}\,[\gamma\cos(\theta/2) - 2\omega_a'\sin(\theta/2)]\}$$
$$+\tfrac{1}{2}i\{-2\omega_a'\tau + (1/\omega_p)\sqrt{2\omega_a'\varepsilon}\,[\gamma\sin(\theta/2) + 2\omega_a'\cos(\theta/2)]\}. \quad (31)$$

As $\varepsilon \to 0$, the real part $\varphi(\omega)$ of the exponent stays around $-\gamma\tau/2$, and its imaginary part $\psi(\omega)$ remains close to $-\omega_a'\tau$. Thus, the only stationary-phase contour that passes through $\omega = \omega_a$ is the one associated with the (constant) amplitude $-\gamma\tau/2$, and the only steepest-descent path that goes through $\omega = \omega_a$ is the one whose (constant) phase is $-\omega'_a\tau$.

Finally, in the vicinity of the pole $\omega = \omega_b$ of the refractive index, we have

$$i\omega[n(\omega) - \tau] \;\to\; i\omega_b\left[\frac{(\omega_b - \omega_a)^{\frac{1}{2}}(\omega_b + \omega_a^*)^{\frac{1}{2}}}{(\omega - \omega_b)^{\frac{1}{2}}(\omega_b + \omega_b^*)^{\frac{1}{2}}} - \tau\right]. \tag{32}$$

Writing $\omega - \omega_b = \varepsilon\exp(i\theta)$, where $|\varepsilon| \ll 1$, and $0 \le \theta < 2\pi$ is measured counterclockwise from the positive real axis $\omega'$, we find

$$i\omega[n(\omega) - \tau] \;\to\; (-\omega''_b + i\omega'_b)\left[\frac{i(\omega'_a - \omega'_b)^{\frac{1}{2}}(\omega'_a + \omega'_b)^{\frac{1}{2}}}{\sqrt{\varepsilon}\exp(i\theta/2)(2\omega'_b)^{\frac{1}{2}}} - \tau\right]$$

$$= (\tfrac{1}{2}\gamma + i\omega'_b)\left\{\sqrt{(\omega'^2_a - \omega'^2_b)/(2\omega'_b\varepsilon)}\,[\sin(\theta/2) + i\cos(\theta/2)] - \tau\right\}$$

$$= -\left\{(\omega_p/\sqrt{2\omega'_b\varepsilon})[\omega'_b\cos(\theta/2) - \tfrac{1}{2}\gamma\sin(\theta/2)] + \tfrac{1}{2}\gamma\tau\right\}$$

$$+ i\left\{(\omega_p/\sqrt{2\omega'_b\varepsilon})[\omega'_b\sin(\theta/2) + \tfrac{1}{2}\gamma\cos(\theta/2)] - \omega'_b\tau\right\}. \tag{33}$$

If we now fix $\varepsilon$ and travel around the pole (by allowing $\theta$ to climb from 0 to $2\pi$), Eq.(33) reveals that $\varphi(\omega)$ rises from $-(\omega_p\sqrt{\omega'_b/2\varepsilon} + \tfrac{1}{2}\gamma\tau)$ to a peak at $(\omega_p\omega_r/\sqrt{2\omega'_b\varepsilon} - \tfrac{1}{2}\gamma\tau)$, then declines again to acquire its final value of $(\omega_p\sqrt{\omega'_b/2\varepsilon} - \tfrac{1}{2}\gamma\tau)$; see Fig.8. Similarly, $\psi(\omega)$ starts at $(\tfrac{1}{2}\omega_p\gamma/\sqrt{2\omega'_b\varepsilon} - \omega'_b\tau)$, rises to a peak value of $(\omega_p\omega_r/\sqrt{2\omega'_b\varepsilon} - \omega'_b\tau)$, then declines steadily to arrive at its final value of $-(\tfrac{1}{2}\omega_p\gamma/\sqrt{2\omega'_b\varepsilon} + \omega'_b\tau)$. Thus, for sufficiently small $\varepsilon$, we find a large range of positive as well as negative values for both $\varphi(\omega)$ and $\psi(\omega)$ in the immediate vicinity of the pole at $\omega_b$. All contours of constant amplitude (i.e., stationary-phase), and also all contours of constant phase (i.e., steepest-descent) must thus spiral their way into this pole, unless they are terminated at the adjacent branch-cut. In other words, the natural tendency of the steepest-descent trajectories is to ultimately get absorbed by the pole in the direction of $\theta = 2\pi - \tan^{-1}(\gamma/2\omega'_b)$, while that of the stationary-phase trajectories is to get absorbed by the same pole in the direction of $\theta = \pi - \tan^{-1}(\gamma/2\omega'_b)$.

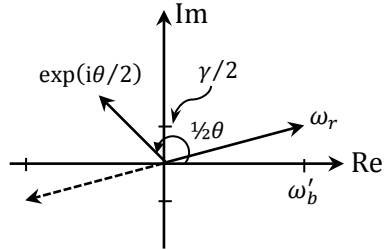

**Fig.8**. Multiplication of $(\omega'_b + i\gamma/2)$ into $\exp(i\theta/2) = \cos(\theta/2) + i\sin(\theta/2)$ is tantamount to a counterclockwise rotation of the vector of length $\omega_r = \sqrt{\omega'^2_b + (\gamma/2)^2}$ through the angle $\theta/2$. The projection of the rotated vector onto the horizontal axis will then be $\omega'_b\cos(\theta/2) - \tfrac{1}{2}\gamma\sin(\theta/2)$. Similarly, the projection of the vector onto the vertical axis will be $\omega'_b\sin(\theta/2) + \tfrac{1}{2}\gamma\cos(\theta/2)$.

Figure 9 shows a set of steepest-descent contours for the parameter set $\omega_p = 14\omega_{\text{ref}}$, $\omega_r = 4\omega_{\text{ref}}$, and $\gamma = 0.1\omega_{\text{ref}}$, computed at $\tau = 1.1$. The central ellipse represents a saddle-point contour corresponding to $\psi_0 = 0$. The contours inside this large ellipse on the right-hand side represent positive values of $\psi_0$, whereas those inside and on the left correspond to negative $\psi_0$. Outside the ellipse, the contours on the right belong to negative $\psi_0$, while those on the left represent positive $\psi_0$.

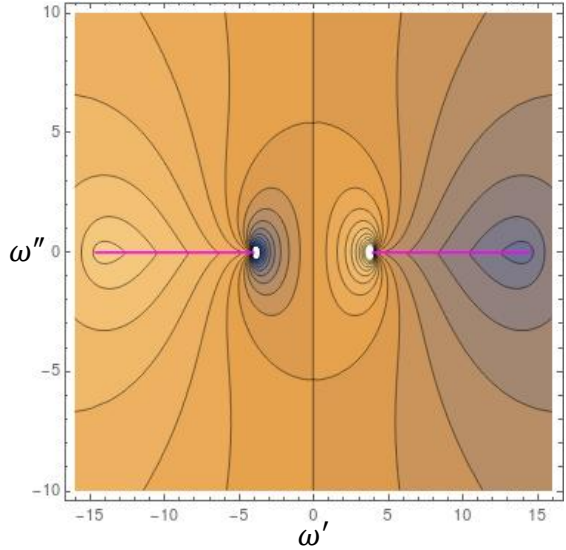

**Fig.9**. Steepest-descent contours in the $\omega$-plane, obtained by setting $\psi(\omega) = \psi_0$ for different values of the constant $\psi_0$. The purple lines are the branch-cuts, located below the real axis at $\omega'' = -\gamma/2$. The branch-cut on the right extends from $\omega_b$ to $\omega_a$, its mirror image forming the branch-cut on the left-hand side.

Careful examination of the real part $\varphi(\omega)$ of the exponent of the integrand in Eq.(1) reveals some important features of the steepest-descent contours. For example, on the real axis, where $\omega'' = 0$, $\varphi(\omega)$ is always negative, since $\omega'n''(\omega)$ is always positive. That the phase of $n(\omega)$ is positive for $\omega > 0$, and negative for $\omega < 0$, can be inferred from the triangle shown in Fig.10.

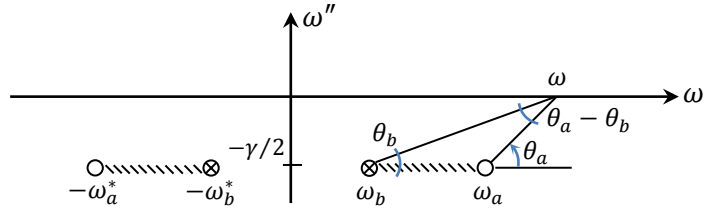

**Fig.10**. The refractive index $n(\omega)$ is determined by the square root of the product of distances from $\omega$ to the zeros at $\omega_a$ and $-\omega_a^*$, divided by the product of distances from $\omega$ to the poles at $\omega_b$ and $-\omega_b^*$. The phase of $n(\omega)$ is ½$(\theta_a - \theta_b)$ plus the corresponding contribution from the pole-zero pair located on the opposite side of the imaginary axis. When $\omega$ is on the real axis, the phase of $n(\omega)$ is positive for $\omega > 0$, and negative for $\omega < 0$. The phase of $n(\omega)$ vanishes when $\omega$ moves onto the imaginary axis.

On the imaginary axis, where $\omega' = 0$ and $n(\omega)$ is always real and greater than 1.0, we have $\psi(\omega'') = 0$ and $\varphi(\omega'') = -[n'(\omega'') - \tau]\omega''$. Since $n'(\omega'')$ is at its maximum at $\omega'' = -\gamma/2$, where $n(\omega) = \omega_a'/\omega_b'$, it is seen that, moving up the positive $\omega''$-axis, the sign of $\varphi(\omega'')$ switches from negative to positive at the point where $n'(\omega'') = \tau$, if such a point is ever reached. A corresponding change of sign occurs on the negative $\omega''$-axis. Note, however, that $\varphi(\omega'')$ has an additional change of sign at the origin, where $\omega''$ switches sign.

Moving along the $\omega''$-axis, considering that $\mathrm{d}\varphi/\mathrm{d}\omega'' = -\dot{n}'(\omega'')\omega'' - [n'(\omega'') - \tau] = 0$ at a saddle-point, we will have $[n'(\omega'') - \tau] = -\dot{n}'(\omega'')\omega''$ and, therefore, $\varphi(\omega'') = \dot{n}'(\omega'')\omega''^2$. When $\omega'' > -\gamma/2$, the derivative of $n(\omega)$ along the $\omega''$-axis is negative and, consequently, $\varphi(\omega)$ at the upper saddle will be negative. The opposite occurs at the lower saddle point.

**6. Finding the saddle points**. The saddle points of the integrand in Eq.(1) are the solutions of the following equation:

$$\tfrac{\mathrm{d}}{\mathrm{d}\omega}[\omega n(\omega) - \omega\tau] = 0 \quad \rightarrow \quad \omega\dot{n}(\omega) + n(\omega) = \tau. \tag{34}$$

The derivative $\dot{n}(\omega)$ of the refractive index $n(\omega)$ with respect to $\omega$ can be directly evaluated from Eq.(2); see Appendix B. Upon substituting the expression for $\dot{n}(\omega)$ into Eq.(34), we find

$$\frac{\omega_p^2(\omega + \tfrac{1}{2}\mathrm{i}\gamma)\omega}{(\omega-\omega_a)(\omega+\omega_a^*)(\omega-\omega_b)(\omega+\omega_b^*)} = \frac{\tau}{n(\omega)} - 1. \tag{35}$$

The solutions of Eq.(35) always appear in pairs (i.e., as mirror images in the imaginary axis)—unless, of course, they already reside on the vertical axis. Being reducible to an 8$^{\text{th}}$ order polynomial equation, Eq.(35) is expected to have eight solutions in the $\omega$-plane, which act as saddle points for the integral in Eq.(1). However, some of the solutions may be overlapping, and some may be unacceptable—because Eq.(35) needs to be squared before becoming a polynomial equation. All in all, Eq.(35) yields a total of four saddle points for the integral in Eq.(1). For values of $\tau$ slightly greater than 1, one saddle resides on the far right-hand side of the $\omega$-plane, sitting slightly below the real axis at $\omega_s \cong [\omega_p/\sqrt{2(\tau-1)}] - \mathrm{i}\gamma$, its mirror image being on the far left-hand side; see Appendix C. In this regime of $\tau \gtrsim 1$, the other two saddles reside on the imaginary axis and relatively far from the origin, one on the positive side, the other on the negative side of the $\omega''$-axis. As $\tau$ increases, the pair of saddle-points on the far right and far left move toward the zeros of the refractive index at $\omega = \omega_a$ and $\omega = -\omega_a^*$, always remaining below the real axis. The two saddle-points on the $\omega''$-axis also move toward each other and eventually meet at a point on the imaginary axis, just below the origin at $\omega \cong -\mathrm{i}\gamma/3$. Further increases in $\tau$ cause the latter pair of saddles to branch out and move away from the $\omega''$-axis, one moving to the right, the other to the left of the imaginary axis. Figure 11 shows typical trajectories of the four saddle points in the special case when $\omega_p = 14\omega_{\text{ref}}$, $\omega_r = 4\omega_{\text{ref}}$, and $\gamma = 0.1\omega_{\text{ref}}$, with $\tau$ starting slightly above 1.0, then rising to large values. Note in both Figs.11(a) and 11(b) that the vertical axes are greatly magnified relative to the horizontal axes.

The solution of Eq.(34) for upper and lower saddles can be obtained from a graph such as that shown in Fig.12; see Appendix D for the relevant analysis. When $\tau$ is only slightly above 1.0, there will be two crossing points, one above and the other below the origin. With the passage of time (i.e., as $\tau$ increases), there comes a point in time when $\tau$ equals $n(0) = \sqrt{1 + (\omega_p/\omega_r)^2}$. At this point, Eq.(35) is exactly satisfied, and $\omega_{s3} = 0$ becomes a saddle point. (It is this instant in time that Brillouin[1] has identified as the beginning of the second forerunner; see Fig.1(b).) A short time later, one arrives at $\tau = n(-\tfrac{1}{2}\mathrm{i}\gamma) = \sqrt{1 + \omega_p^2/(\omega_r^2 - \tfrac{1}{4}\gamma^2)}$, where, once again, Eq.(35) is exactly satisfied, with the corresponding saddle at $\omega_{s4} = -\tfrac{1}{2}\mathrm{i}\gamma$. The temporal evolution of these saddle points, i.e., $\omega_{s3}$ joining $\omega_{s4}$, then branching out into the $\omega'\omega''$-plane, as depicted in Fig.11(b), can be qualitatively understood by examining Fig.12; their quantitative evaluation, however, requires detailed numerical analysis.

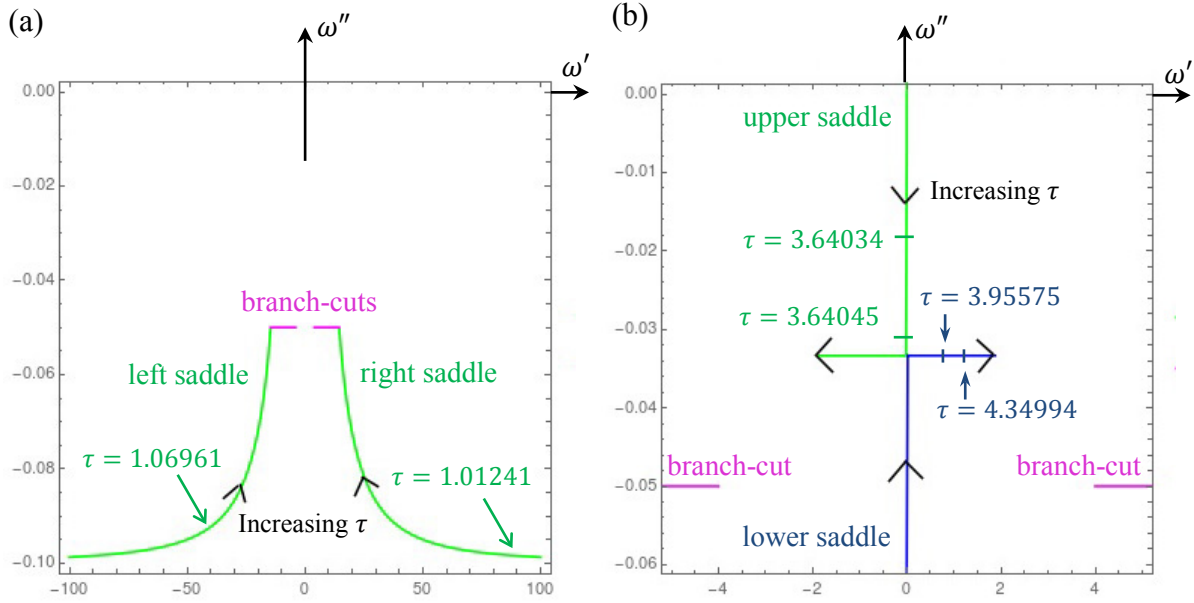

**Fig.11**. (a) Trajectories of the left and right saddles for $\tau$ ranging from 1.01 to 101. For $1.0 \leq \tau \lesssim 1.01$, the locations of these saddle points are well approximated by $\omega_{\text{saddle}} \cong \pm \omega_p/\sqrt{2(\tau-1)} - i\gamma$. (b) Trajectories of the upper and lower saddles on and near the imaginary axis for $\tau$ ranging from 3.63 to 4.75. These saddle points, which are initially on the imaginary axis and relatively far from the origin, move slowly at first, as $\tau$ rises above 1.0; they then accelerate toward their meeting point at $\omega \cong -i\gamma/3$ and split apart, moving into the third and fourth quadrants of the $\omega$-plane.

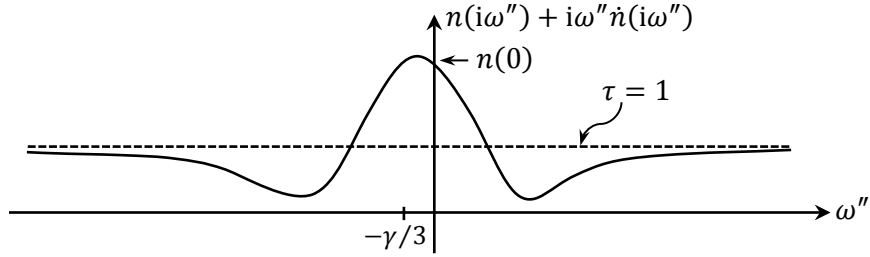

**Fig.12**. Typical plot of the function $n(i\omega'') + i\omega''\dot{n}(i\omega'')$ along the $\omega''$-axis. The peak appears at $\omega'' \cong -\gamma/3$, where the value of the function is generally greater than 1.0. The peak is flanked by two minima, beyond which the function rises to its asymptotic value of 1.0 as $|\omega''| \to \infty$. When $\tau$ initially exceeds 1.0, there are two crossing points corresponding to two saddles on the imaginary axis. Eventually, however, these crossing points merge at $\omega'' \cong -\gamma/3$, whence they branch out into the 3rd and 4th quadrants of the $\omega$-plane. In the opposite direction, when $\tau$ drops below 1.0, there will be four crossing points at first, placing all four saddles on the imaginary axis. As $\tau$ declines toward the bottom of the curve, the saddles merge pairwise, below which they move into the right and left halves of the $\omega$-plane.

According to Eq.(34), the saddle points are located where $n(\omega) + \omega\dot{n}(\omega) = \tau$. This requires that the real and imaginary parts of the equation be simultaneously satisfied. Considering that $\dot{n}(\omega)$ can be equivalently written as $\partial n(\omega)/\partial \omega'$, we will have

$$n''(\omega) + \omega'[\partial n''(\omega)/\partial \omega'] + \omega''[\partial n'(\omega)/\partial \omega'] = 0, \qquad (36a)$$

$$n'(\omega) + \omega'[\partial n'(\omega)/\partial \omega'] - \omega''[\partial n''(\omega)/\partial \omega'] = \tau. \qquad (36b)$$

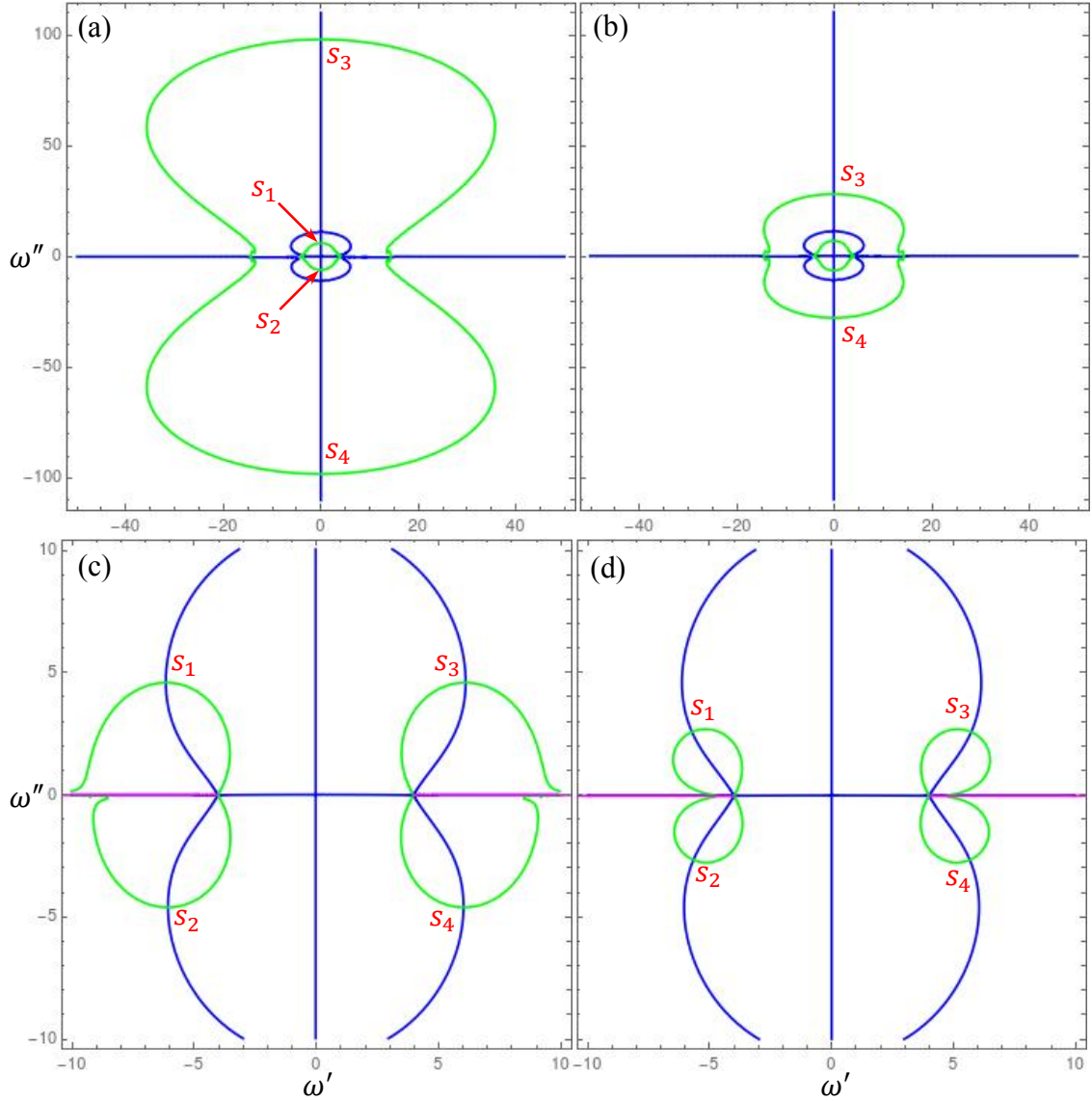

**Fig.13**. Evolution of the saddle points as $\tau$ drops further and further below 1.0. The four saddles are at the various intersections of the contours where $\partial \varphi / \partial \omega' = 0$ (blue) and $\partial \psi / \partial \omega' = 0$ (green). The blue curves, which are independent of $\tau$, include the $\omega''$-axis as well as a curve very close to (but not coincident with) the $\omega'$-axis. The two saddles on the $\omega''$-axis that are closest to the origin, denoted by $s_1, s_2$, are more or less stationary so long as the other two saddles on the $\omega''$-axis, identified as $s_3, s_4$, stay away from them. (Whereas $s_1, s_2$ are the continuation of the upper and lower saddles of the $\tau > 1$ regime, $s_3, s_4$ are the high-frequency saddles in the case of $\tau > 1$, which have jumped over to the $\omega''$-axis and are now moving toward $s_1, s_2$ as $\tau$ continues to decrease.) When $s_3$ and $s_4$ eventually collide with $s_1$ and $s_2$, all four saddle points leave the $\omega''$-axis and move into the $\omega$-plane. As $\tau$ continues its decline (eventually into negative territory), all four saddles approach the poles $(\omega_b, -\omega_b^*)$ of the refractive index along the (fixed) blue curves.

The first of the above equations is independent of $\tau$, specifying curves in the $\omega$-plane over which the function on the left-hand side of Eq.(36a) is zero. Figure 13 shows the contours over which $\partial \varphi / \partial \omega' = 0$ (blue) and $\partial \psi / \partial \omega' = 0$ (green) at several instants of time where $\tau \leq 1$. (The case of $\tau \leq 1$ is of significance for bandlimited incident signals, even though it had no relevance in the original Sommerfeld-Brillouin analysis.) In frames (a) and (b) of Fig.13, all four saddles ($s_1, s_2, s_3, s_4$) are located on the $\omega''$-axis. As time continues to decline, these saddles merge (pairwise) and branch out into the $\omega$-plane, always maintaining a symmetric position with respect to the imaginary axis.

The blue curves in Fig.13 include the entire imaginary axis, along which $\omega' = 0$, $n''(\omega) = 0$, and $n'(\omega)$ does not vary with slight changes in $\omega'$. Another one of the blue curves appears to overlap the real axis, but this is just because the scale of the graph is coarse. In reality, this blue curve is quite close, but not everywhere parallel, to the real axis, as confirmed by the trajectory of the saddle points depicted in Fig.11(a). The green curves in Fig.13 are plots of Eq.(36b) for various values of $\tau$, where, in all four frames, $\tau < 1.0$. As $\tau$ declines (eventually becoming negative), the green curves shrink, and the saddle points $s_3$ and $s_4$ move toward the other two saddle points ($s_1, s_2$). The four saddles then merge (pairwise) and move away from the imaginary axis and into the $\omega$-plane, to be eventually absorbed by the poles at $\omega_b$ and $-\omega_b^*$.

**7. Steepest-descent contours for integrating bandlimited signals.** Figure 14 shows the two steepest-descent contours (red and dark blue) that form a closed loop with the integration interval ($\omega_{min}, \omega_{max}$) over which the incident waveform has a nonzero spectral distribution (light blue). The branch-cuts (in purple) are shown slightly below the real axis. The observed signal arriving at $z = z_o$ is computed by adding the contributions of the two steepest-descent contours that originate at the end-points $\omega_{min}$ and $\omega_{max}$ of the incident spectrum, and terminate on the adjacent branch-cut. In the magnified view of the relevant contours in Fig.14(b), the arrows indicate the direction of decline of $\varphi(\omega)$. Since we have chosen $\omega_{max} < \omega_b'$ in this example, there is no need for additional contours to close the integration path. For sufficiently large $z_o$, the observed signal is well approximated by evaluating the integrals in the vicinity of $\omega_{min}$ and $\omega_{max}$ along the corresponding steepest-descent trajectories, as the integrand declines exponentially and rapidly away from these end-points.

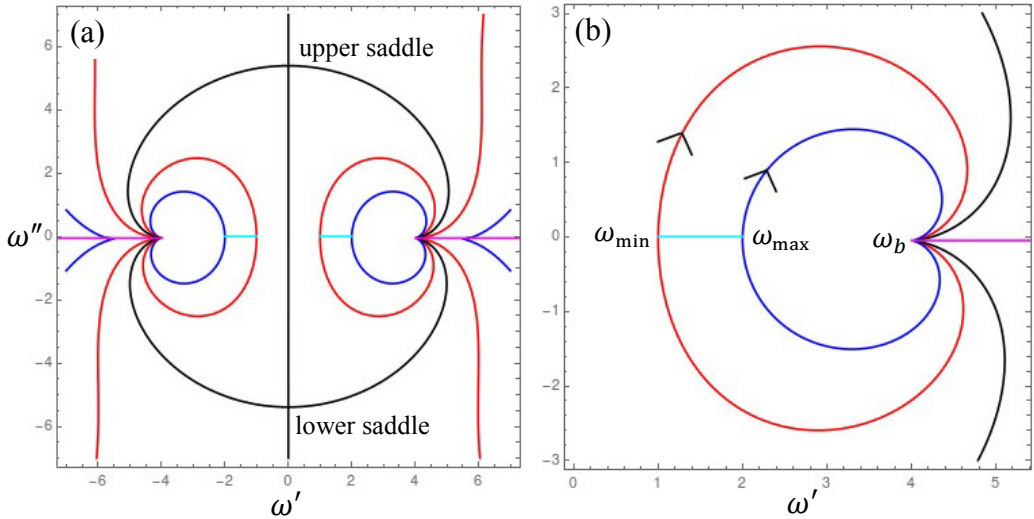

**Fig.14**. (a) Steepest-descent contours in the $\omega$-plane corresponding to $\omega_p = 14\omega_{ref}$, $\omega_r = 4\omega_{ref}$, and $\gamma = 0.1\omega_{ref}$ at $\tau = 1.01$. The branch-cuts are in purple, and the integration intervals $\omega_{min} \leq |\omega| \leq \omega_{max}$ are depicted in light blue. (b) Magnified view in the vicinity of the integration interval ($\omega_{min}, \omega_{max}$), where two steepest-descent contours begin at the end-points of the integration interval and terminate on the adjacent branch-cut. The arrows indicate the direction in which $\varphi(\omega)$ declines.

The situation is fundamentally different in the case depicted in Fig.15, where $\tau = 1.05$ and $\omega_{max}$ is chosen to be much greater than $\omega'_a$ (i.e., the real part of the zero $\omega_a$ of the refractive index). Here, the end-point contours (red and dark blue) do *not* suffice to close the integration path, and a third contour (green) is needed to create a closed loop together with the end-point contours and the straight-line segment (light blue) that defines the spectral range $(\omega_{min}, \omega_{max})$ of the incident waveform. This new steepest-descent contour (green) is chosen to go through a saddle point, which is located slightly below the $\omega'$-axis at the crossing point of the two green curves in Fig.15. The observed signal at $z = z_o$ now has contributions from the end-points as well as the saddle-point. For sufficiently large values of $\zeta = z_o/c$, the observed signal will be well approximated by the sum of the end-point contributions along their respective steepest-descent contours in the vicinity of $\omega_{min}$ and $\omega_{max}$, plus the contribution from the contour that goes through the saddle-point in the vicinity of that saddle point. Considering that the saddle in Fig.15 is located on the left-hand side of $\omega_{max}$, its contribution to the observed signal at $z = z_o$ does *not* constitute a superoscillation, simply because $\omega'_{saddle}$ is within the spectral range of the incident waveform.

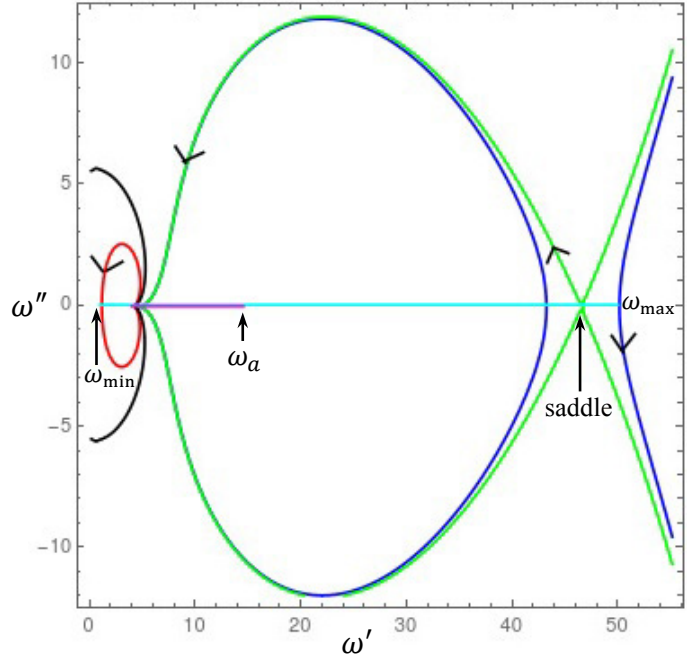

**Fig.15**. Steepest-descent contours in the $\omega$-plane corresponding to $\omega_p = 14\omega_{ref}$, $\omega_r = 4\omega_{ref}$, and $\gamma = 0.1\omega_{ref}$ at $\tau = 1.05$. The integration interval $(\omega_{min}, \omega_{max})$, depicted in light blue, is fairly broad, with $\omega_{max} > \omega'_a$. A dark blue contour emerges from the end-point $\omega_{max}$ and goes to infinity, whence up comes a green contour that passes through a saddle-point on its way to termination on the branch-cut (purple). A red contour then starts on the branch-cut and reaches the end-point $\omega_{min}$ of the spectral range. The arrows indicate the direction that must be travelled along each contour to obtain a closed integration path. While this happens to be the direction of decline of $\varphi(\omega)$ for the blue curve and also for the green curve (beyond the saddle), for the red curve it is the direction along which $\varphi(\omega)$ increases.

For a specific numerical example, consider the incident spectrum $\mathcal{E}(\omega) = \sin^{100}(\pi\omega/2\omega_s)$ shown in Fig.16(a), where $|\omega| \leq 2\omega_s$. The host medium's refractive index in the present example is assumed to have $\omega_p = \omega_r = 4\omega_{ref}$ and $\gamma = 0.1\omega_{ref}$. With these choices, the branch-cuts (shown in purple) are slightly below the real axis (at $\omega''/\omega_{ref} = -0.05$), occupying the intervals $[-5.66, -4.00]$ and $[4.00, 5.66]$ along the normalized real axis $\omega'/\omega_{ref}$. The center frequency for the incident spectrum is chosen in the middle of the branch-cut at $\omega_s = 4.83\omega_{ref}$, and the observation point is located at $\zeta = z_0/c = 5000/\omega_{ref}$.

Figure 16(b) shows the various steepest-descent contours at $\tau = 1.4$, where the high-frequency saddles are at $\omega_{saddle} = (\pm 7.83 - 0.07i)\omega_{ref}$, the saddle on the positive imaginary axis is at $\omega'' = 0.44\omega_{ref}$, and the end-points are at $\pm\omega_{max} = \pm 2\omega_s = \pm 9.66\omega_{ref}$. The numerical value of $\mathcal{E}(\omega)$ at the point on the real axis just above the high-frequency saddle is $8.9 \times 10^{-26}$. The exact value of $E(z_0, t)$, obtained by direct numerical integration of Eq.(1) from $-2\omega_s$ to $2\omega_s$ along the light blue line in Fig.16(b), is found to be $-2.073 \times 10^{-91}$. An approximate value for $E(z_0, t)$ can also be obtained by integrating along the five steepest-descent contours shown in Fig.16(b). The main contribution, at $-2.365 \times 10^{-91}$, comes from the two contours that pass through the left and right saddle-points. The contour through the upper saddle makes the relatively minor contribution of $6.837 \times 10^{-95}$, and the end-point contours together contribute only $\sim 10^{-11312}$, which is totally negligible. It is clear that the high-frequency saddles completely dominate the signal at the observation point.

It is thus seen that high-frequency forerunners can also exist in the case of bandlimited incident beams. However, the oscillation frequency of these forerunners is always within the bandwidth of the incident signal, which overrules the possibility of superoscillations.

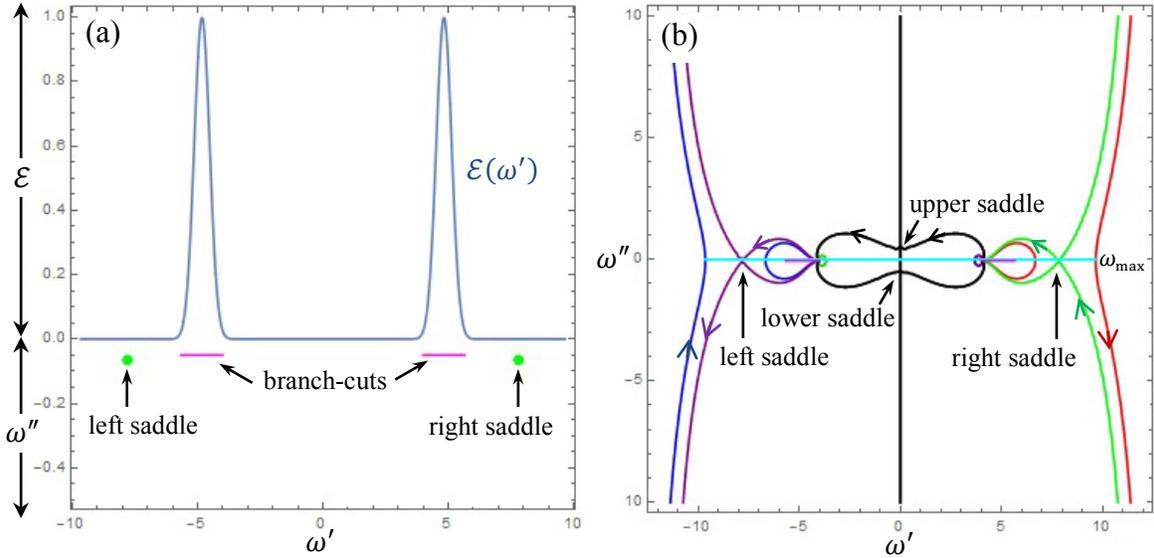

**Fig.16**. Evaluation of the end-point and saddle-point contributions to the observed signal at $z = z_0$ for the bandlimited incident spectral distribution $\mathcal{E}(\omega) = \sin^{100}(\pi\omega/2\omega_s)$, where $|\omega| \leq 2\omega_s$. (a) Spectral profile, together with the placing of the branch-cuts (purple) and high-frequency saddles (green dots) in the $\omega$-plane. The end-points are at $\pm 2\omega_s/\omega_{ref} = \pm 9.66$. (b) Constellation of steepest-decent contours that form a closed loop with the integration interval $(-\omega_{max}, \omega_{max}) = (-2\omega_s, 2\omega_s)$. From the end-point on the right-hand side, the red contour brings the path to infinity, where the green contour takes over and brings it back (through the right saddle) onto the branch-cut at right. From here, one moves along the black curve (through the upper saddle) onto the branch-cut at left. Subsequently, the purple contour brings the integration path from the left branch-cut to infinity (through the left saddle), where the blue curve takes over and finally closes the path on the left end-point at $\omega = -\omega_{max}$.

**8. Concluding remarks.** The Sommerfeld-Brillouin forerunners (or precursors) consist of the *actual* frequencies that are present in the spectrum of the incident light pulse, but are systematically squeezed into the short time interval $(z_0/c, z_0/v_g)$, where $z_0$ is the distance to the observation point, $c$ is the speed of light in vacuum, and $v_g$ is the group velocity of propagation at the frequency $\omega_s$ of the incident signal. By demonstrating that the Sommerfeld precursor carries the same energy as is present in the high-frequency tails of the incident spectrum (once absorption has been accounted for), we have ruled out the possibility of this precursor being a form of leaky superoscillation.[10]

A bandlimited light pulse can similarly exhibit high-frequency oscillations at an observation point $z = z_0$ deep inside a dispersive medium. The frequencies of these oscillations, however, do not exceed the highest frequency that is already present within the spectrum of the incident light pulse. These weak, high-frequency oscillations do not appear to be of the leaky type either, as they only emerge when the corresponding saddle-point sweeps below the maximum frequency $\omega_{max}$ of the incident light pulse.

In this paper, we have described a general approach to analyzing the propagation of bandlimited signals deep inside a homogeneous, isotropic, absorptive, and dispersive dielectric medium. The method involves the identification of a number of steepest-descent contours that are subsequently used to replace the original integration path along the spectral bandwidth of the incident signal. The methods of saddle-point and/or stationary-phase approximation can then be used to asymptotically evaluate the resulting integrals along the identified steepest-descent trajectories.

We examined a limited number of situations and addressed a few issues in the context of the theory of superoscillations, but many interesting problems remain that can benefit from the methodology developed in the preceding sections. The super smooth spectra described in Sec.4, Eqs.(26) and (27), for example, pose challenging analytical as well as numerical hurdles in conjunction with their essential singularities at the spectral end-points ($\omega_{min}, \omega_{max}$). It will also be interesting to study situations involving the four saddle-points depicted in Fig.13, since, in the temporal regime of $\tau \leq 1$, a bandlimited incident signal could exhibit behavior that is either dominated by its end-point contours or by the low-frequency saddles of Fig.13. Another subject for future investigation would involve incident spectra that reside in the "twilight zone" that is the boundary between regions of normal and anomalous dispersion.

Finally, the study of gain media, in which the incident wave is amplified (as opposed to being attenuated) might reveal additional interesting features of the precursor signal. In the Lorentz oscillator model of a gain medium, the negative oscillator strength yields the refractive index as $n(\omega) = \sqrt{1 - \omega_p^2/(\omega_r^2 - \omega^2 - i\gamma\omega)}$; here, we have absorbed the magnitude of the oscillator strength into the plasma frequency $\omega_p$. While the poles of this $n(\omega)$ continue to be at $\pm\sqrt{\omega_r^2 - (\gamma/2)^2} - i(\gamma/2)$, its zeros now shift to $\pm\sqrt{\omega_r^2 - \omega_p^2 - (\gamma/2)^2} - i(\gamma/2)$. An examination of the branch-cuts of this $n(\omega)$ reveals that $\omega_p$ must remain below $\omega_r$, lest a portion of the branch-cut creeps into the upper half of the $\omega$-plane, a situation that is prohibited by the relativistic requirement that no signal should arrive at $z = z_0$ prior to $t = z_0/c$. This observation provides a physical justification as to why the real and imaginary parts of $n(\omega)$ are required to obey the Kramers-Kronig relations.[2,9] For a more nuanced treatment of gain media—one that argues in favor of extending the range of $\omega_p$ beyond $\omega_r$ without violating the tenets of special relativity—the reader is referred to the works of J. Skaar.[11]

## Appendix A

At the location of the right-hand saddle, that is, at $\omega_{\text{saddle}} \cong [\omega_p/\sqrt{2(\tau-1)}] - i\gamma$, assuming that $\tau \gtrsim 1$, the exponent of the integrand in Eq.(1) can be approximated as follows:

$$i\zeta\omega_{\text{saddle}}[n(\omega_{\text{saddle}}) - \tau] \cong i\zeta\left(\frac{\omega_p}{\sqrt{2(\tau-1)}} - i\gamma\right)\left(\sqrt{1 + \frac{\omega_p^2}{\omega_r^2 - \omega_{\text{saddle}}^2 - i\gamma\omega_{\text{saddle}}}} - \tau\right)$$

$$\cong \zeta\left(\gamma + \frac{i\omega_p}{\sqrt{2(\tau-1)}}\right)\left(\sqrt{1 + \frac{\omega_p^2}{\underbrace{\omega_r^2 - [\omega_p^2/2(\tau-1)] + 2i\gamma[\omega_p/\sqrt{2(\tau-1)}] + \gamma^2 - i\gamma[\omega_p/\sqrt{2(\tau-1)}] - \gamma^2}_{\cong 0}}} - \tau\right)$$

$$\cong \zeta\left(\gamma + \frac{i\omega_p}{\sqrt{2(\tau-1)}}\right)\left(\sqrt{1 - \frac{2(\tau-1)}{1 - i\gamma[\sqrt{2(\tau-1)}/\omega_p]}} - \tau\right)$$

$$\cong \zeta\left(\gamma + \frac{i\omega_p}{\sqrt{2(\tau-1)}}\right)\left[1 - (\tau-1)\left(1 + \frac{i\gamma\sqrt{2(\tau-1)}}{\omega_p}\right) - \tau\right]$$

$$= -\zeta(\tau-1)\left(\gamma + \frac{i\omega_p}{\sqrt{2(\tau-1)}}\right)\left(2 + \frac{i\gamma\sqrt{2(\tau-1)}}{\omega_p}\right)$$

$$= -\zeta(\tau-1)\left(\gamma + \frac{2i\omega_p}{\sqrt{2(\tau-1)}} + \underbrace{\frac{i\gamma^2\sqrt{2(\tau-1)}}{\omega_p}}_{\cong 0}\right)$$

$$\cong -\gamma(t - z_o/c) - i\omega_p\sqrt{2(z_o/c)(t - z_o/c)}. \tag{A1}$$

The real part of the above expression is the exponent of the damping factor introduced by Brillouin[1,4] as a correction to Sommerfeld's precursor amplitude—since Sommerfeld's original derivation required that $\gamma$ be set to zero. The imaginary part of the exponent in Eq.(A1) eventually becomes the chirped frequency of oscillations that appears in Eq.(10).

Brillouin's damping factor is precisely what one would obtain by allowing the frequency content of the incident spectrum at $\omega = \omega_{\text{spc}} = \omega_p/\sqrt{2(\tau-1)}$ to propagate directly to the observation point at $z = z_o$. To see this, note that the real part of the exponent of the propagation factor at $\omega = \omega_{\text{spc}}$ is given by

$$\text{Re}[i\zeta\omega_{\text{spc}}n(\omega_{\text{spc}})] = -\zeta\omega_{\text{spc}}n''(\omega_{\text{spc}}) = -\zeta\omega_{\text{spc}}\,\text{Im}\left(\sqrt{1 + \frac{\omega_p^2}{\omega_r^2 - \omega_{\text{spc}}^2 - i\gamma\omega_{\text{spc}}}}\right)$$

$$\cong -\zeta\omega_{\text{spc}}\,\text{Im}\left[1 - \frac{\omega_p^2}{2(\omega_{\text{spc}}^2 + i\gamma\omega_{\text{spc}})}\right] = \tfrac{1}{2}\zeta\,\text{Im}\left(\frac{\omega_p^2}{\omega_{\text{spc}} + i\gamma}\right)$$

$$\cong \tfrac{1}{2}\zeta\,\text{Im}\left[\frac{\omega_p^2}{\omega_{\text{spc}}}\left(1 - \frac{i\gamma}{\omega_{\text{spc}}}\right)\right] = -\tfrac{1}{2}\gamma\zeta(\omega_p/\omega_{\text{spc}})^2$$

$$= -\gamma\zeta(\tau-1) = -\gamma(t - z_o/c). \tag{A2}$$

The equivalence of the damping factor obtained by Brillouin's saddle-point approximation of the integral in Eq.(1), and that obtained by direct propagation of the incident spectral content at $\omega = \omega_{\text{spc}}$ (albeit at the corresponding group velocity $v_g(\omega_{\text{spc}})$), is further confirmation that the Sommerfeld precursor is *not* a manifestation of superoscillatory behavior, but rather a systematic temporal arrangement of the frequencies that are already present in the incident waveform.

## Appendix B

The derivative with respect to $\omega$ of the refractive index $n(\omega)$ of Eq.(2) is evaluated as follows:

$$\dot{n}(\omega) = \frac{dn(\omega)}{d\omega} = \frac{d}{d\omega}[(\omega - \omega_a)^{\frac{1}{2}}(\omega + \omega_a^*)^{\frac{1}{2}}(\omega - \omega_b)^{-\frac{1}{2}}(\omega + \omega_b^*)^{-\frac{1}{2}}]$$

$$= \frac{1}{2}(\omega - \omega_a)^{-\frac{1}{2}}(\omega + \omega_a^*)^{-\frac{1}{2}}(\omega - \omega_b)^{-\frac{1}{2}}(\omega + \omega_b^*)^{-\frac{1}{2}}$$
$$\times [(\omega + \omega_a^*) + (\omega - \omega_a) - (\omega - \omega_a)(\omega + \omega_a^*)(\omega - \omega_b)^{-1} - (\omega - \omega_a)(\omega + \omega_a^*)(\omega + \omega_b^*)^{-1}]$$

$$= \frac{1}{2}[(\omega - \omega_a)(\omega + \omega_a^*)(\omega - \omega_b)(\omega + \omega_b^*)]^{-\frac{1}{2}}$$
$$\times \left[2\omega + \omega_a^* - \omega_a - \frac{(\omega - \omega_a)(\omega + \omega_a^*)}{(\omega - \omega_b)(\omega + \omega_b^*)}(2\omega + \omega_b^* - \omega_b)\right]$$

$$= (\omega + \tfrac{1}{2}i\gamma)\left[1 - \frac{(\omega - \omega_a)(\omega + \omega_a^*)}{(\omega - \omega_b)(\omega + \omega_b^*)}\right][(\omega - \omega_a)(\omega + \omega_a^*)(\omega - \omega_b)(\omega + \omega_b^*)]^{-\frac{1}{2}}$$

$$= \omega_p^2(\omega + \tfrac{1}{2}i\gamma)(\omega - \omega_a)^{-1/2}(\omega + \omega_a^*)^{-1/2}(\omega - \omega_b)^{-3/2}(\omega + \omega_b^*)^{-3/2}$$

$$= \frac{\omega_p^2(\omega + \tfrac{1}{2}i\gamma)n(\omega)}{(\omega - \omega_a)(\omega + \omega_a^*)(\omega - \omega_b)(\omega + \omega_b^*)}. \tag{B1}$$

Note that, if $\omega$ happens to be on the imaginary axis, $n(\omega)$ will be purely real, but $\dot{n}(\omega)$ will be purely imaginary, due to the fact that the denominator of $dn(\omega)/d\omega$ will be imaginary.

## Appendix C

For large values of $\omega$, Eq.(35) can be solved with the aid of elementary approximation methods, as follows:

$$\omega\dot{n}(\omega) + n(\omega) = \tau$$

$$\rightarrow \frac{\omega_p^2(\omega + i\gamma/2)\omega}{(\omega - \omega_a)^{1/2}(\omega + \omega_a^*)^{1/2}(\omega - \omega_b)^{3/2}(\omega + \omega_b^*)^{3/2}} + \sqrt{1 + \frac{\omega_p^2}{\omega_r^2 - \omega^2 - i\gamma\omega}} = \tau$$

$$\rightarrow \frac{(\omega_p/\omega)^2(1 + i\gamma/2\omega)}{(1 - \omega_a/\omega)^{1/2}(1 + \omega_a^*/\omega)^{1/2}(1 - \omega_b/\omega)^{3/2}(1 + \omega_b^*/\omega)^{3/2}} + 1 - \frac{(\omega_p/\omega)^2}{2[1 + (i\gamma/\omega) - (\omega_r/\omega)^2]} \cong \tau$$

$$\rightarrow (\omega_p/\omega)^2(1 + i\gamma/2\omega)(1 + \omega_a/2\omega)(1 - \omega_a^*/2\omega)(1 + 3\omega_b/2\omega)(1 - 3\omega_b^*/2\omega)$$
$$- \tfrac{1}{2}(\omega_p/\omega)^2[1 - (i\gamma/\omega) + (\omega_r/\omega)^2] \cong \tau - 1$$

$$\rightarrow (\omega_p/\omega)^2[1 + (i\gamma/2\omega) + \underbrace{(\omega_a - \omega_a^*)/2\omega}_{-i\gamma} + \underbrace{3(\omega_b - \omega_b^*)/2\omega}_{-i\gamma} - \tfrac{1}{2} + \tfrac{1}{2}(i\gamma/\omega)] \cong \tau - 1$$

$$\rightarrow (\omega_p/\omega)^2[\tfrac{1}{2} - i(\gamma/\omega)] \cong \tau - 1$$

$$\rightarrow \omega/\omega_p \cong \pm\sqrt{1 - i(2\gamma/\omega)}/\sqrt{2(\tau - 1)}$$

$$\rightarrow \omega \cong \pm\frac{\omega_p}{\sqrt{2(\tau - 1)}}[1 - i(\gamma/\omega)] \qquad \rightarrow \qquad \omega_{\text{saddle}} \cong \pm\frac{\omega_p}{\sqrt{2(\tau - 1)}} - i\gamma. \tag{C1}$$

As $t$ rises beyond $z_0/c$, the parameter $\tau$ goes above 1.0, and the right and left saddles at $\omega_{\text{saddle}}$ of Eq.(C1) move rather swiftly from $\pm\infty - i\gamma$ to $\sim \pm 10\omega_p - i\gamma$ (at $\tau = 1.005$), to $\sim \pm 5\omega_p - i\gamma$ (at $\tau = 1.02$), and then to $\sim \pm 3\omega_p - i\gamma$ (at $\tau = 1.055$). Needless to say, as $\tau$ drifts further and further beyond 1.0, the approximations that lead to Eq.(C1) make the above estimate of $\omega_{\text{saddle}}$ less and less trustworthy.

## Appendix D

The saddle-points on the imaginary axis can be found by substituting $i\omega''$ for $\omega$, then writing

$$n(i\omega'') = \left(1 + \frac{\omega_p^2}{\omega''^2 + \gamma\omega'' + \omega_r^2}\right)^{1/2}. \tag{D1}$$

$$\dot{n}(i\omega'') = \frac{i\omega_p^2(2\omega'' + \gamma)}{2(\omega''^2 + \gamma\omega'' + \omega_r^2)^2}\left(1 + \frac{\omega_p^2}{\omega''^2 + \gamma\omega'' + \omega_r^2}\right)^{-1/2}. \tag{D2}$$

$$n(i\omega'') + i\omega''\dot{n}(i\omega'') = \left(1 + \frac{\omega_p^2}{\omega''^2 + \gamma\omega'' + \omega_r^2}\right)^{1/2} - \frac{\omega_p^2(2\omega'' + \gamma)\omega''}{2(\omega''^2 + \gamma\omega'' + \omega_r^2)^2}\left(1 + \frac{\omega_p^2}{\omega''^2 + \gamma\omega'' + \omega_r^2}\right)^{-1/2} = \tau. \tag{D3}$$

For any given value of $\tau$, Eq.(D3) must be solved numerically to reveal the location of the saddle-point(s) on the $\omega''$-axis. To better understand the nature of these solutions, we equate the derivative with respect to $\omega''$ of the left-hand side of Eq.(D3) to zero, arriving at

$$\frac{d[n(i\omega'') + i\omega''\dot{n}(i\omega'')]}{d\omega''} = \omega_p^2(\omega''^2 + \gamma\omega'' + \omega_r^2)^{-5/2}(\omega''^2 + \gamma\omega'' + \omega_r^2 + \omega_p^2)^{-3/2}$$
$$\times \left[\omega''(2\omega'' + \gamma)^2(\omega''^2 + \gamma\omega'' + \omega_r^2 + \tfrac{3}{4}\omega_p^2)\right.$$
$$\left. -(3\omega'' + \gamma)(\omega''^2 + \gamma\omega'' + \omega_r^2)(\omega''^2 + \gamma\omega'' + \omega_r^2 + \omega_p^2)\right] = 0. \tag{D4}$$

For typical values of the parameter set $(\omega_p, \omega_r, \gamma)$, the 5$^{\text{th}}$ order polynomial equation appearing in Eq.(D4) has five solutions of which only three are real-valued. These represent the locations of a single maximum and two nearby minima of the function on the left-hand side of Eq.(D3). The location of the maximum, inferred from Eq.(D4), is $\omega'' \cong -\gamma/3$. An examination of Eqs.(D3) and (D4) reveals the general profile of $n(i\omega'') + i\omega''\dot{n}(i\omega'')$ as depicted in Fig.12.

**References**

1. Léon Brillouin, *Wave Propagation and Group Velocity*, Academic Press, New York (1960).
2. J. D. Jackson, *Classical Electrodynamics*, 2$^{\text{nd}}$ edition, Sec.7.11, Wiley, New York (1975); see also 3$^{\text{rd}}$ edition, Sec. 7.11, Wiley, New York (1999).
3. Arnold Sommerfeld, "*Über die Fortpflanzung des Lichtes in dispergierenden Medien*," Annalen der Physik (Leipzig) **44**, 177-202 (1914) − English translation appearing in Léon Brillouin, *Wave Propagation and Group Velocity*, Academic Press, New York (1960).
4. Léon Brillouin, "*Über die Fortpflanzung des Lichtes in dispergierenden Medien*," Annalen der Physik (Leipzig) **44**, 203-240 (1914).
5. Arnold Sommerfeld, in "*Festschrift zum 70. Geburtstag von Heinrich Weber*," pp 338-374, Teubner, Leipzig (1912).
6. F. B. Hildebrand, *Advanced Calculus for Applications*, 2$^{\text{nd}}$ edition, Prentice-Hall, New Jersey (1976).
7. J. Mathews and R. L. Walker, *Mathematical Methods for Physics*, 2$^{\text{nd}}$ edition, Benjamin/Cummings Publishing, California (1970).
8. C. M. Bender and S. A. Orszag, *Advanced Mathematical Methods for Scientists and Engineers I: Asymptotic Methods and Perturbation Theory*, Springer -Verlag, New York (1999).
9. M. Mansuripur, *Field, Force, Energy and Momentum in Classical Electrodynamics* (Revised Edition), Bentham Science Publishers, Sharjah, UAE (2017).
10. M.V. Berry, "*Superoscillations and leaky spectra*," J. Phys. A: Math. Theor. 52, 015202 (2019).
11. J. Skaar, "Fresnel equations and the refractive index of active media," Physical Review E **73**, 026605 (2006).